\documentclass[twocolumn,english,aps,reprint,prd,showpacs,floatfix]{revtex4}
\usepackage[T1]{fontenc}
\usepackage[latin1]{inputenc}
\usepackage{array}
\usepackage{prettyref}
\usepackage{amsmath}
\usepackage{graphicx}
\usepackage{amssymb}

\makeatletter

\providecommand{\tabularnewline}{\\}

\usepackage{units}

\def\nat{Nature\ }
\def\aap{Astron.\ Astrophys.\ }
\def\apj{Astrophys.\ J.\ }
\def\apjl{Astrophys.\ J.\ Lett.\ }
\def\apjs{Astrophys.\ J.\ Supp.\ }

\def\mnras{Mon.\ Not.\ Roy.\ Astron.\ Soc.\ }

\usepackage{babel}
\makeatother
\begin{document}

\preprint{The preprint version\#}

\title{On the Detectability of Galactic Dark Matter Annihilation \linebreak
into Monochromatic Gamma-rays}

\author{Zhi-Cheng TANG}

\email{tangzhch@ihep.ac.cn}

\author{Qiang YUAN}

\author{Xiao-Jun BI}

\author{Guo-Ming CHEN}

\affiliation{Institute of High Energy Physics, Chinese Academy of Sciences}

\begin{abstract}
Monochromatic $\gamma$-rays are thought to be the smoking gun signal
for identifying the dark matter annihilation. However, the flux of
monochromatic $\gamma$-rays is usually suppressed by the virtual
quantum effects since dark matter should be neutral and does not couple
with $\gamma$-rays directly. In the work we study the detection strategy
of the monochromatic $\gamma$-rays in a future space-based detector.
The monochromatic $\gamma$-ray flux is calculated by assuming supersymmetric
neutralino as a typical dark matter candidate. We discuss both the
detection focusing on the Galactic center and in a scan mode which
detects $\gamma$-rays from the whole Galactic halo are compared.
The detector performance for the purpose of monochromatic $\gamma$-rays
detection, with different energy and angular resolution, field of
view, background rejection efficiencies, is carefully studied with
both analytical and fast Monte-Carlo method. 
\end{abstract}


\maketitle

\section{Introduction}

The existence of dark matter (DM) in the Universe is widely accepted
nowadays. The evidences come from many astronomical observations,
which observed the gravitational effects of dark matter in different
spatial scales, from dwarf galaxies, galaxies, galaxy clusters to
the cosmological scale. It is recognized that DM particle should be
neutral, cold and non-baryonic, which can only exist in theories beyond
the standard model of particle physics. Among the large amount of
DM candidates proposed in the literature, the weakly interacting massive
particles (WIMP) are the most favored one, which can account for the
observed DM density naturally.

In order to determine the nature of the WIMP DM particles, we generally
have three ways to probe the interaction between DM particles and
the standard particles: the direct detection measures the the scattering
by DM with the detector nuclei; or produce DM particles directly in
a powerful collider such as LHC or ILC; finally the indirect detection
searches for the DM annihilation or decay products in cosmic rays
(CRs), including gamma-rays, electrons, positrons, protons, antiprotons
and neutrinos. For indirect detection the gamma-rays are usually the
best probe of DM because they are not deflected by the magnetic field
during the propagation. Further the technology of high energy gamma
ray detector was developed very fast in the last years. In general
the gamma-rays are produced through the hadronization and decay of
the DM annihilation/decay final states, and have continuous energy
spectra. Since there are large diffuse backgrounds of gamma-rays,
it is usually not easy to figure out the DM signals from the background.
Another annihilation channel is to monochromatic gamma-rays with small
branching ratio via loop diagram. The monochromatic gamma-rays by
DM annihilation are usually taken as the smoking-gun of the DM signal
since there are no such background from astrophysical processes. If
the detector has very good energy resolution, the background will
be suppressed and the detectability will be improved.

In this work we discuss DM annihilation into monochromatic photons,
$\chi_{0}\chi_{0}\rightarrow\gamma\gamma$ and $\chi_{0}\chi_{0}\rightarrow\gamma Z_{0}$,
with energy of the gamma-ray photon $m_{\chi}$ and $m_{\chi}-m_{Z}^{2}/4m_{\chi}$
respectively. Here we neglect the kinetic energy of DM particles since
its movement is non-relativistic today. For very massive DM $m_{\chi}\gg m_{Z}$,
the photon energy of the two channels are identical and can not be
distinguished in experiments. This work tries to give the perspective
of detecting such line spectrum gamma-rays from DM annihilation in
the Milky Way, and show the requirements for detector design.

The $\gamma$-ray flux from DM annihilation is proportional to the
annihilation cross section and DM density square. As a typical WIMP
DM we consider the lightest neutralino in the minimal supersymmetric
standard model (MSSM) as an explicit example in our calculation. The
cross section can be computed given the MSSM model parameters. In
this work we will employ DarkSUSY package \cite{2004JCAP...07..008G}
to scan the MSSM parameter space. As to the Galactic DM density profile,
numerical simulations indicate that DM is highly concentrated in the
halo center. Therefore the Galactic center (GC) is usually the first
choice searching for DM signals. In addition, there are also a large
amount of substructures existing in the halo, mostly in the outer
part of the halo. The contribution from substructures will also be
discussed. The backgrounds for the monochromatic photon detection
include CR nuclei (mainly protons and Helium for energies $\lesssim10$
TeV), electrons, diffuse continuous $\gamma$-rays and the $\gamma$-ray
point sources in the GC region. The nuclei and electrons can be rejected
through the particle identification technique of the detector. For
the diffuse $\gamma$-ray background we need a high energy resolution
to suppress the background.

The paper is organized as follow. The MSSM model and the Galactic
DM density distribution will be presented in Sec. \ref{sec:input}.
In Sec. \ref{sec:detectability} the possible backgrounds in our study
are introduced, especially those in the GC region.  Sec. \ref{sec:analytical}
gives the results of detection sensitivity for the GC region through
simple analytical estimate. We show in Sec. \ref{sec:mc} the sensitivity
for all-sky observation in the scan mode, with substructures included,
using Monte-Carlo simulation. Finally we draw the conclusion and discussion
in Sec. \ref{sec:discuss}.

\section{Dark matter annihilation into monochromatic gamma-rays \label{sec:input}}

The $\gamma$-ray flux from DM annihilation can be written as \begin{equation}
\phi(\psi)=\frac{\rho_{\odot}^{2}R_{\odot}}{4\pi}\times\frac{N\langle\sigma v\rangle}{2m_{\chi}^{2}}\times J(\psi),\label{eq:flux}\end{equation}
 where $\psi$ is a specified direction away from the Galactic center,
$\rho_{\odot}\approx0.4$ GeV cm$^{-3}$ \cite{2009arXiv0907.0018C,2010arXiv1003.3101S}
is the local DM density, $R_{\odot}\approx8.5$ kpc is the distance
from the Earth to the GC, $m_{\chi}$ is the mass of DM particle,
$\langle\sigma v\rangle$ is the velocity weighted thermal average
annihilation cross section, the multiplicity $N=1,\,2$ for $\gamma Z_{0}$
and $\gamma\gamma$ channels respectively. Finally $J(\psi)$ is the
line-of-sight integral of the density square $J(\psi)=\frac{1}{\rho_{\odot}^{2}R_{\odot}}\int\rho^{2}(l)dl$.
The $\gamma$-ray flux depends on both the particle parameters and
the density distribution of DM.

\subsection{MSSM dark matter model\label{sec:MSSM-dark-matter}}

Without losing the generality we take neutralino in MSSM as a typical
WIMP DM in this work. In order to reduce the number of free parameters
in MSSM we only take a few relevant parameters to our discussion free,
as done in Ref. \cite{2004JCAP...07..008G}, that is, \begin{equation}
\mu,\, M_{2},\,\, M_{1},\,\tan\beta,\, M_{A},\, m_{0},\, A_{b},\, A_{t},\end{equation}
 where $\mu$ is the Higgsino mass parameter, $M_{2}$ and $M_{1}$
are the wino and bino mass parameters respectively, $\tan\beta$ is
the ratio of the vacuum expectations of the two Higgs fields, $M_{A}$
is the mass of pseudo-scalar Higgs boson, $m_{0}$ is the universal
sfermion mass, $A_{b}$ and $A_{t}$ are the trilinear soft breaking
parameters and the corresponding parameters for the first two generations
are assumed zero.

We employ DarkSUSY to explore the parameter space of the phenomenological
MSSM model \cite{2004JCAP...07..008G}. The scan ranges of these parameters
are: $50$ GeV $<|\mu|,M_{2},M_{1},M_{A},m_{0}<$ $10$ TeV, $1.1<\tan\beta<55$,
${\textrm{sign}}(\mu)=\pm1$, $-3m_{0}<A_{t},A_{b}<3m_{0}$. There
are also other constraints from the theoretical consistency requirements
and the accelerator data. Finally, we require the relic density of
DM to be $\Omega_{\chi}h^{2}<0.128$ according to the $3\sigma$ upper
limits of WMAP seven year results \cite{2010arXiv1001.4538K}.

\subsection{Density distribution}

The most precise knowledge of the density profile of DM inside the
halo comes from numerical simulations. Navarro et al. (1997) found
that the density profile is universal for halos of different scales,
referred as Navarro-Frenk-White (NFW) profile \cite{1997ApJ...490..493N}\begin{equation}
\rho(r)=\frac{\rho_{s}}{\left(r/r_{s}\right)\left(1+r/r_{s}\right)^{2}},\end{equation}
 where $\rho_{s}$ and $r_{s}$ are two scale parameters depending
on the mass and concentration of the halo. However, due to the limit
of resolution, some other density profiles with different central
behavior were also proposed in literature. For example Moore et al.
(1999) proposed the density profile with a much steeper inner slope
\cite{1999MNRAS.310.1147M}\begin{equation}
\rho(r)=\frac{\rho_{s}}{\left(r/r_{s}\right)^{1.5}\left[1+\left(r/r_{s}\right)^{1.5}\right]}.\end{equation}
 The recent simulations tended to favor the Einasto profile with a
gradual flattening of the logarithm slope of the inner behavior \cite{2010MNRAS.402...21N}.
There were also studies showing that the density profile might be
non-universal \cite{2000ApJ...529L..69J}. Considering the diversity
of the inner profile of DM density distribution, we adopt NFW and
Moore profiles for this study. The model parameters are $r_{s}=20$
kpc, $\rho_{s}=0.35$ GeV cm$^{-3}$ for NFW, and $r_{s}=28$ kpc,
$\rho_{s}=0.078$ GeV cm$^{-3}$ for Moore profile respectively. The
local density for these parameter sets is $0.4$ GeV cm$^{-3}$.

To avoid the divergence of density when $r\rightarrow0$, a cutoff
scale is applied considering the fact that there should be a balance
between the gravitational infall and the annihilation \cite{1992PhLB..294..221B}.
For common parameters of DM particle the maximum density is estimated
to be $\sim10^{18}$ M$_{\odot}$ kpc$^{-3}$ \cite{2008A&A...479..427L}.

\subsection{Substructures}

The cosmological structures form hierarchically in the cold dark matter
scenario, that is, the DM collapses to form small halos first, then
grows to larger and larger halos through accretion and merger. Numerical
simulations show that there are a large number of subhalos surviving
the merger history and existing in the Milky Way dark matter halo
\cite{1998MNRAS.299..728T,1999ApJ...524L..19M,2000ApJ...544..616G,2003ApJ...598...49Z}.
The dwarf galaxies are a part of the Galactic subhalos which have
been observed.

According to the simulations, the number density distribution of subhalos
as a function of its mass and location can be parameterized as \cite{2004MNRAS.355..819G,2004MNRAS.352..535D}\begin{equation}
\frac{dn}{dm_{\textrm{sub}}\cdot4\pi r^{2}dr}=\frac{n_{0}}{1+(r/r_{h})^{2}}\times\left(\frac{m_{\textrm{sub}}}{m_{\textrm{host}}}\right)^{-\alpha},\end{equation}
 with $r_{h}\approx0.14r_{\textrm{host}}$ for galaxy scale halo \cite{2004MNRAS.352..535D}
and $\alpha\approx1.9$ \cite{2004MNRAS.355..819G,2008MNRAS.391.1685S}.
The normalization $n_{0}$ is fixed by setting $N(>10^{8}\,{\textrm{M}}_{\odot})\approx100$
(see \cite{2008A&A...479..427L} and references therein). The density
profile of each subhalo is also assumed to be NFW or Moore profile.
To determine the density parameters of each subhalo, we use the concentration-mass
relation given in Ref. \cite{2001MNRAS.321..559B}. The procedure
of determining the parameters is as follows. For a halo with mass
$m_{\textrm{sub}}$, the concentration is derived according to the
$c_{\textrm{sub}}-m_{\textrm{sub}}$ relation. Then we have $r_{s}^{\textrm{NFW}}=r_{\textrm{sub}}/c_{\textrm{sub}}$
and $r_{s}^{\textrm{Moore}}=r_{\textrm{sub}}/0.63c_{\textrm{sub}}$
following the definition of concentration \cite{2001MNRAS.321..559B}.
Finally $\rho_{s}$ is determined by the subhalo mass.

To calculate the annihilation flux of photons from the subhalo population,
we define the average density square as \begin{equation}
\langle\rho_{\textrm{sub}}^{2}\rangle(r)=\int dm_{\textrm{sub}}\frac{dn}{dm_{\textrm{sub}}\cdot4\pi r^{2}dr}\times\int_{V_{\textrm{sub}}}\rho_{\textrm{sub}}^{2}dV,\end{equation}
 where $dV$ integrates over the volume of the subhalo. Replacing
$\rho^{2}$ in $J(\psi)$ in Eq. (\ref{eq:flux}) with $\langle\rho_{\textrm{sub}}^{2}\rangle$
we calculate the photon flux from DM annihilation in the subhalos.

\begin{figure}[hbt]
\includegraphics[%
  width=0.49\textwidth]{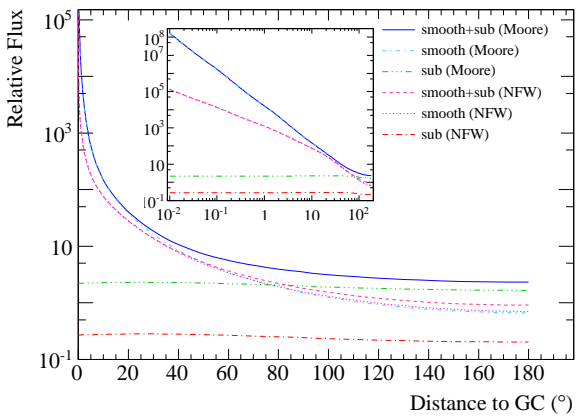}

\caption{\label{fig:dm-dist}Angular distribution of DM annihilation luminosity
($J$ factor) in the Milky Way. The inner plot uses log scale for
$x$ axis to show the details in the most central region.}
\end{figure}

The $J(\psi)$ factor as a function of the angle $\psi$ away from
the GC direction is shown in Fig. \ref{fig:dm-dist}. It can be seen
that for the smooth halo the annihilation flux is highly concentrated
in the GC. The substructure contribution is nearly isotropic in all
directions. At large angles away from the GC the substructure component
may dominate the annihilation flux for the Moore profile.

\section{Detectability analysis\label{sec:detectability}}

\subsection{Backgrounds\label{sec:Backgrounds}}

We first introduce the backgrounds which are essential for the detectability
analysis of the DM-induced monochromatic $\gamma$-ray signal. The
backgrounds include charged CR particles, such as all kinds of nuclei,
electrons and positrons, and continuous $\gamma$-rays.

The CR nuclei and electrons/positrons can be rejected through the
design of particle identification technique of the detector, however,
there still are a few particles which may be misidentified as photons
and form the background. The combined nuclei flux, mostly proton and
Helium nuclei, can be written as \cite{2001ICRC....5.1643G}\begin{equation}
\phi_{n}(E)=1.49\left(\frac{E}{\textrm{GeV}}\right)^{-2.74}{\textrm{cm}}^{-2}{\textrm{s}}^{-1}{\textrm{sr}}^{-1}{\textrm{GeV}}^{-1},\end{equation}
 which is an empirical formula from combined result of many measurements.
For the electron plus positron spectrum, we adopt a broken power law
parameterization 

\begin{eqnarray}
\phi_{e}(E) & = & 1.5\times10^{-11}\left[1+\left(\frac{E}{\unit[900]{GeV}}\right)^{10/3}\right]^{-0.33}\nonumber \\
 &  & \times\left(\frac{E}{\unit[900]{GeV}}\right)^{-3.0}\unit{cm^{-2}s^{-1}sr^{-1}GeV^{-1}}\end{eqnarray}
 according to the recent measurements by Fermi \cite{2009PhRvL.102r1101A}
and HESS \cite{2008PhRvL.101z1104A,2009A&A...508..561A}. In the following
we employ two efficiencies, $\eta_{n}$ and $\eta_{e}$, to represent
the rejection power of the charged CRs.

Then we come to the continuous $\gamma$-ray backgrounds. The first
$\gamma$-ray background is the all-sky diffuse $\gamma$-ray emission,
including Galactic and extra-galactic. For the extra-galactic $\gamma$-ray
background we use the new measurement made by Fermi \cite{2010PhRvL.104j1101A}\begin{equation}
\phi_{\gamma}^{\textrm{extra}}(E)=6.57\times10^{-7}\left(\frac{E}{\textrm{GeV}}\right)^{-2.4}{\textrm{cm}}^{-2}{\textrm{s}}^{-1}{\textrm{sr}}^{-1}{\textrm{GeV}}^{-1}.\end{equation}
 The Fermi result of extra-galactic $\gamma$-ray emission is steeper
than that obtained by EGRET \cite{1998ApJ...494..523S}, which will
result in an order of magnitude lower background when extrapolating
to high energies ($\sim$TeV).

Fermi collaboration also reported some data about the Galactic diffuse
$\gamma$-ray emission (e.g., \cite{2009ApJ...703.1249A,2009PhRvL.103y1101A,2010PhRvL.104j1101A}),
which is consistent with the results given by EGRET except the {}``GeV
excess'' \cite{1997ApJ...481..205H}. Since the full Fermi data are
unavailable at the present stage, we use the EGRET data about the
Galactic diffuse $\gamma$-ray emission in this work. The Galactic
diffuse $\gamma$-ray flux is parameterized as \cite{1998APh.....9..137B}\begin{equation}
\phi_{\gamma}^{\textrm{galac}}(E)=N_{0}(l,b)\times10^{-6}\left(\frac{E}{\unit{GeV}}\right)^{-2.7}{\textrm{cm}}^{-2}{\textrm{s}}^{-1}{\textrm{sr}}^{-1}{\textrm{GeV}}^{-1},\label{eq:galac-gamma}\end{equation}
 where \begin{equation}
N_{0}=\left\{ \begin{array}{ll}
\frac{85.5}{\sqrt{1+(l/35)^{2}}\sqrt{1+(b/1.8)^{2}}}+0.5 & |l|\leq30^{\circ}\\
\frac{85.5}{\sqrt{1+(l/35)^{2}}\sqrt{1+\left[b/(1.1+0.022|l|)\right]^{2}}}+0.5 & |l|>30^{\circ}\end{array}\right.,\end{equation}
 in which the galactic longitude $l$ and latitude $b$ are expressed
in unit of degree. To make use of this measurement, we have to extrapolate
Eq. (\ref{eq:galac-gamma}) to higher energies.

Besides the diffuse $\gamma$-ray emission, there are additional sources
in the GC region. Since the GC region is very important for DM searches,
the $\gamma$-ray sources in the GC region are necessary to be paid
more attention. HESS observation showed there was diffuse $\gamma$-ray
emission in the region $|l|<0.8^{\circ},|b|<0.3^{\circ}$ (GC ridge)
on top of the diffuse background \cite{2006Natur.439..695A}. The
spectrum is \begin{equation}
\phi_{\gamma}^{\textrm{GC-diff}}(E)=1.28\times10^{-4}\left(\frac{E}{\textrm{GeV}}\right)^{-2.29}{\textrm{cm}}^{-2}{\textrm{s}}^{-1}{\textrm{sr}}^{-1}{\textrm{GeV}}^{-1}.\end{equation}
 Also there is at least one gamma-ray point source in the Galactic
center area, which is labeled as 3EG J1746-2851 in EGRET catalog \cite{1998A&A...335..161M},
HESS J1745-290 in HESS catalog \cite{2006PhRvL..97v1102A} and 0FGL
J1746.0-2900 in Fermi catalog \cite{2009ApJS..183...46A}. The energy
spectrum of this source given by HESS is \cite{2009A&A...503..817A}\begin{equation}
\phi_{\gamma}^{\textrm{point}}(E)=2.5\times10^{-7}\left(\frac{E}{\textrm{GeV}}\right)^{-2.21}{\textrm{cm}}^{-2}{\textrm{s}}^{-1}{\textrm{GeV}}^{-1},\end{equation}
 which is valid between 200 GeV and 10 TeV.

\begin{figure}[hbt]
\includegraphics[%
  width=0.49\textwidth]{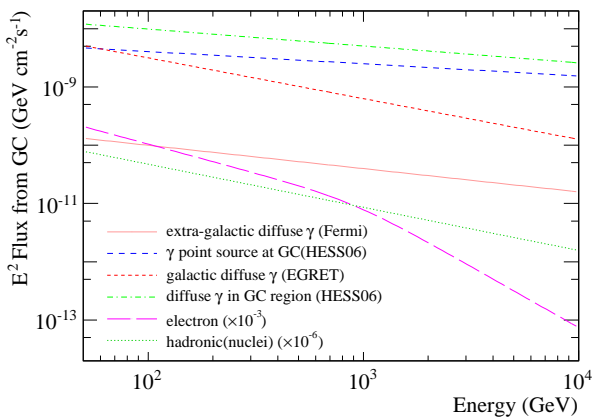}

\caption{Spectra of the possible backgrounds from the 1 degree region around
the GC. \label{fig:bkg-spectrum}}
\end{figure}

The differential fluxes of all the backgrounds mentioned above are
plotted in Fig. \ref{fig:bkg-spectrum}. Here we show the results
within the circle with 1 degree radius around the GC. The electron
and nuclei fluxes are multiplied by factors $10^{-3}$ and $10^{-6}$
respectively, which represent the typical rejection power of a $\gamma$-ray
detector. We can see that the largest background is the diffuse source
and the point source in the GC region. The Galactic diffuse $\gamma$-ray
emission is also important in the GC. However, for the sky regions
far away from the GC, the extra-galactic background would also become
import.

\subsection{Ideal detector}

The sensitivity of monochromatic $\gamma$-ray detection is determined
by the performance of a detector. First of all, we need particle discrimination
to reject most of the charged CRs, say nuclei and electrons/positrons.
This can be done through charge detection, neutron detection and shower
shape identification in calorimeter. The continuous $\gamma$-rays
can be suppressed by improving the detector energy resolution. Furthermore,
as can be seen in Fig. \ref{fig:dm-dist} the small region around
the GC is the best candidate for the line spectrum $\gamma$-rays
detection. Thus good angular resolution will also be effective to
increase the signal-to-noise ratio.

To simplify the study, we characterize the detector with some static
parameters of the performance, that is, the parameters don't change
with incident energy, direction, or particle type, like the energy
resolution, the angular resolution, the field of view, the rejection
power of electrons and nuclei and so on. The effective area of detector
and exposure time are also key factors of the detection, however,
the overall performance can be easily scaled given different effective
area and exposure time. With the results of ideal detectors, we can
provide the requirement of a real detector.

\subsection{Sensitivity calculation}

The event counts on a detector can be written as \begin{equation}
N_{i}=\eta_{i}\cdot T_{\textrm{eff}}\cdot A\cdot f_{i},\label{eq:event}\end{equation}
 where $\eta_{i}$ is the detection efficiency of incident particle,
usually assumed to be $90\%$ for photon, and equal to the rejection
power for electron and nuclei, $T_{\textrm{eff}}$ is the effective
exposure time for a given source or small sky region, $A$ is the
active area of detector, and $f_{i}$ is the flux of signal or background
in given energy range and sky region. For a detector with energy resolution
$\sigma_{e}$, $f_{i}$ can be derived according to the differential
flux $\phi$\begin{equation}
f_{i}=\int_{\Delta\Omega}d\Omega\int_{E_{\gamma}-3\sigma_{E}}^{E_{\gamma}+3\sigma_{E}}dE\,\,\phi_{i}(E,l,b).\end{equation}
 Here we choose the energy window to be $\pm3\sigma_{E}$ around $E_{\gamma}$,
and $\Delta\Omega$ refers to the chosen observation field. Note for
point source the above integral with respect to $\Omega$ disappears.

The detection significance is defined as $S=N_{s}/\sqrt{N_{b}}$ with
$N_{s}$ the count of signal and $N_{b}$ the count of background.
The sensitivity can be derived as the minimal signal flux needed for
a detection significance $S=5$.

\section{GC region sensitivity with analytical estimate \label{sec:analytical}}

In this section we estimate the sensitivity of monochromatic $\gamma$-ray
detection from the GC region by DM annihilation. We choose a typical
planar calorimeter type detector with $1{\textrm{m}}\times1{\textrm{m}}$
size. With one year of flight on the orbit of International Space
Station (ISS) and the assumption of 90 degree field of view, the effective
exposure time is $\sim7.9\times10^{6}{s}$, which is about $1/4$
of one year. More detail of effective exposure time can be found in
Fig. \ref{fig:Effective-survey-time}.

The background event number is \begin{eqnarray}
N_{b} & = & N_{\textrm{galac}}+N_{\textrm{extra}}+N_{e}+N_{n}+N_{\mathrm{point}}+N_{\mathrm{GC-diff}}\nonumber \\
 & = & T_{\textrm{eff}}A\int_{E_{\gamma}-3\sigma_{E}}^{E_{\gamma}+3\sigma_{E}}dE\nonumber \\
 &  & \times\left[\int_{\Delta\Omega}d\Omega\eta\phi_{\gamma}^{\textrm{galac}}(E)\right.\nonumber \\
 &  & +\Delta\Omega\left(\eta\phi_{\gamma}^{\textrm{extra}}(E)+\eta_{e}\phi_{e}(E)+\eta_{n}\phi_{n}(E)\right)\nonumber \\
 &  & +\eta\phi_{\gamma}^{\textrm{point}}(E)+\eta\Omega_{\textrm{GC ridge}}\phi_{\gamma}^{\textrm{GC-diff}}(E)\bigg],\end{eqnarray}
 where $\Delta\Omega=\pi(\sigma_{d}+1)^{2}\times(\pi/180)^{2}$ is
the solid angle considered for event direction selection, in order
to keep the same number of signal events, where $\sigma_{d}$ is the
detector resolution angle in unit of degree. Another method of events
direction selection is to keep events of $1^{\circ}$ from GC after
point spreading of detector angle resolution. Considering that the
largest backgrounds come from point-like sources from GC, and they
expands the same as the cuspy Moore dark matter profile, the one-degree
selection after expansion would result in same level of decreasing
of both signal and background events number, which means poorer sensitivity.
On the other hand, if we select events with larger radius, we can
keep basically the same number of signal events, and better sensitivity.
Therefore the minimal monochromatic photon flux from DM annihilation
within $\sim1$ degree around the GC is $f_{\textrm{min}}=5\sqrt{N_{b}}/\eta T_{\textrm{eff}}A$,
for a $5\sigma$ detection.

\begin{figure*}[hbt]
\includegraphics[%
  width=0.49\textwidth]{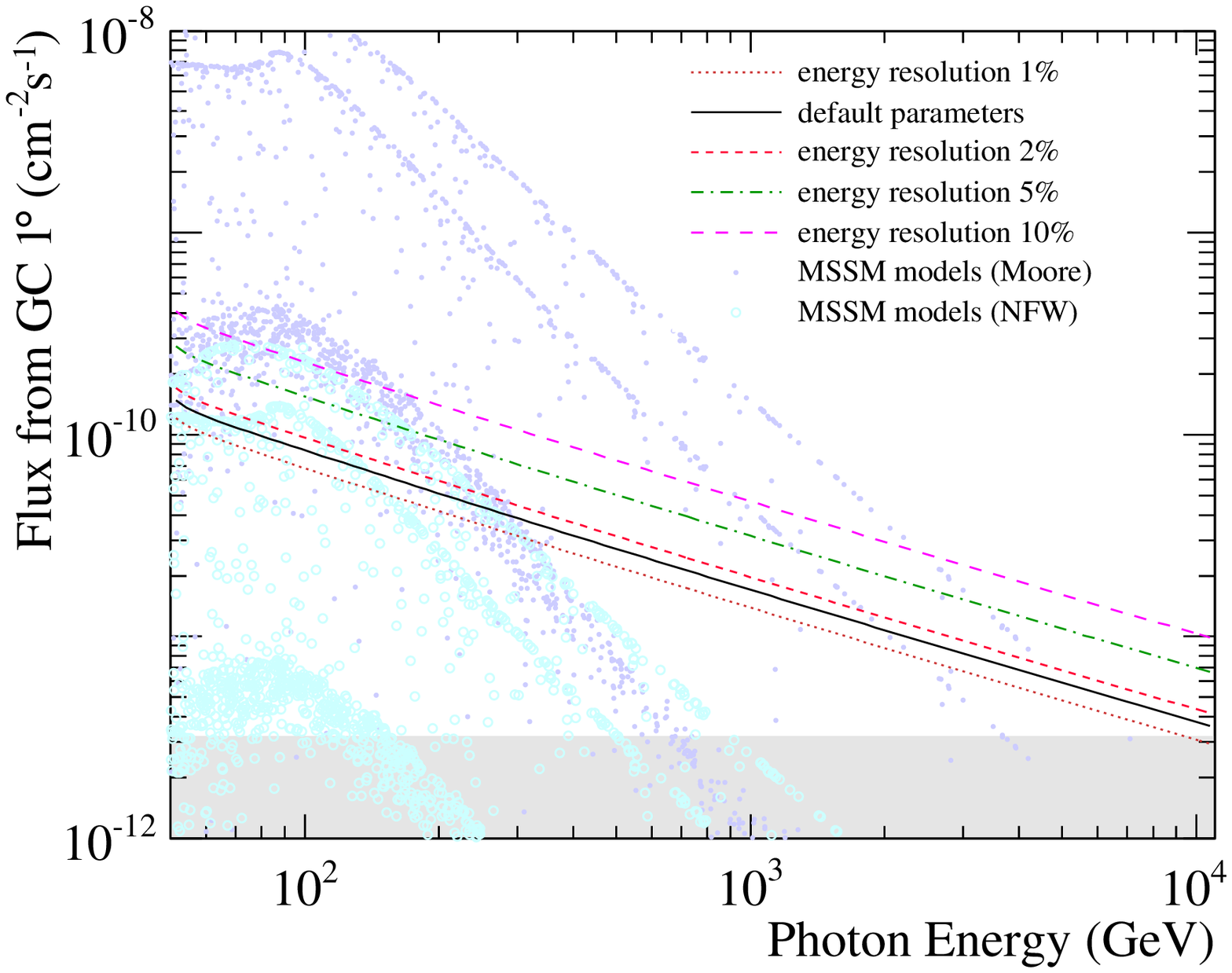} \includegraphics[%
  width=0.49\textwidth]{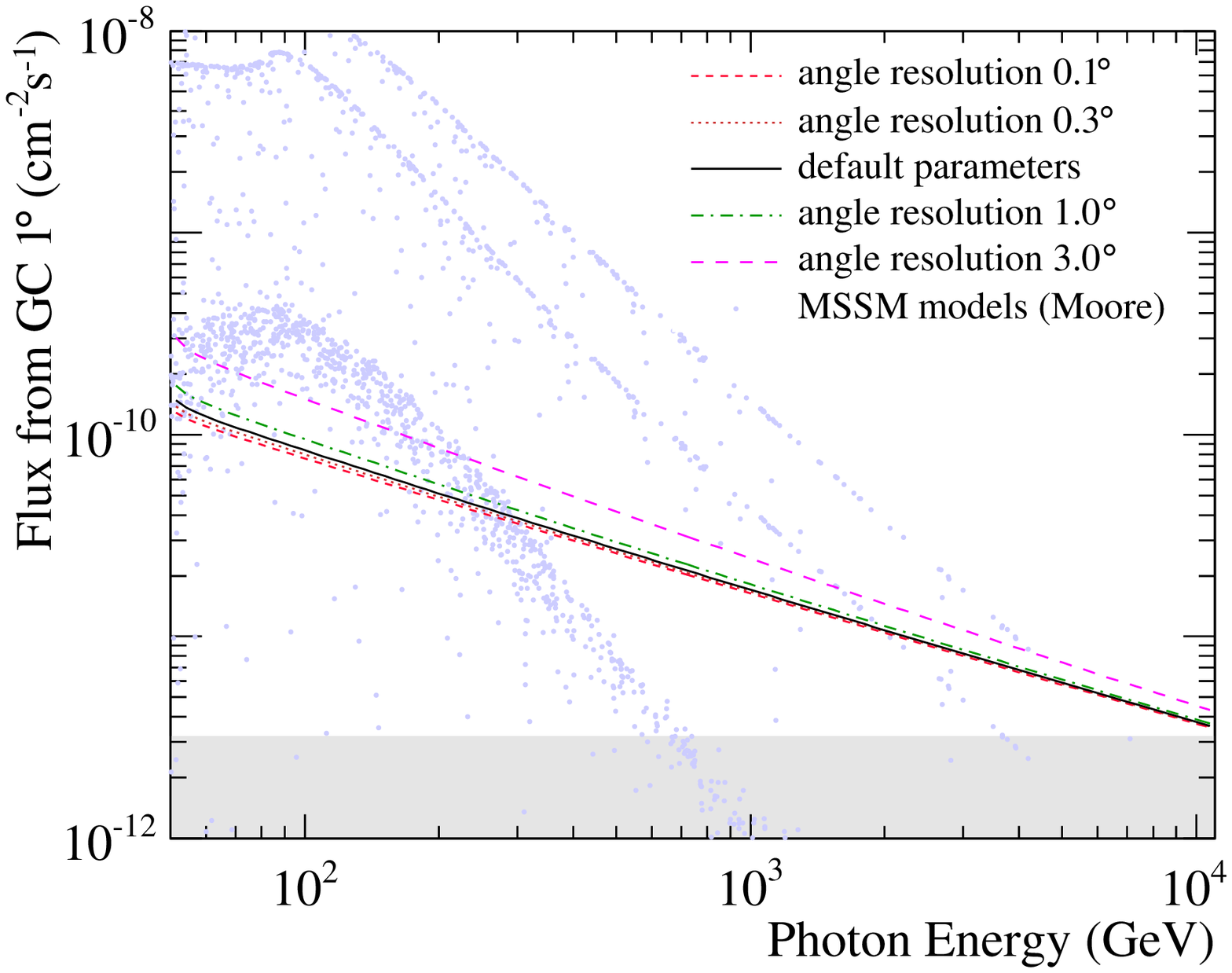} \includegraphics[%
  width=0.49\textwidth]{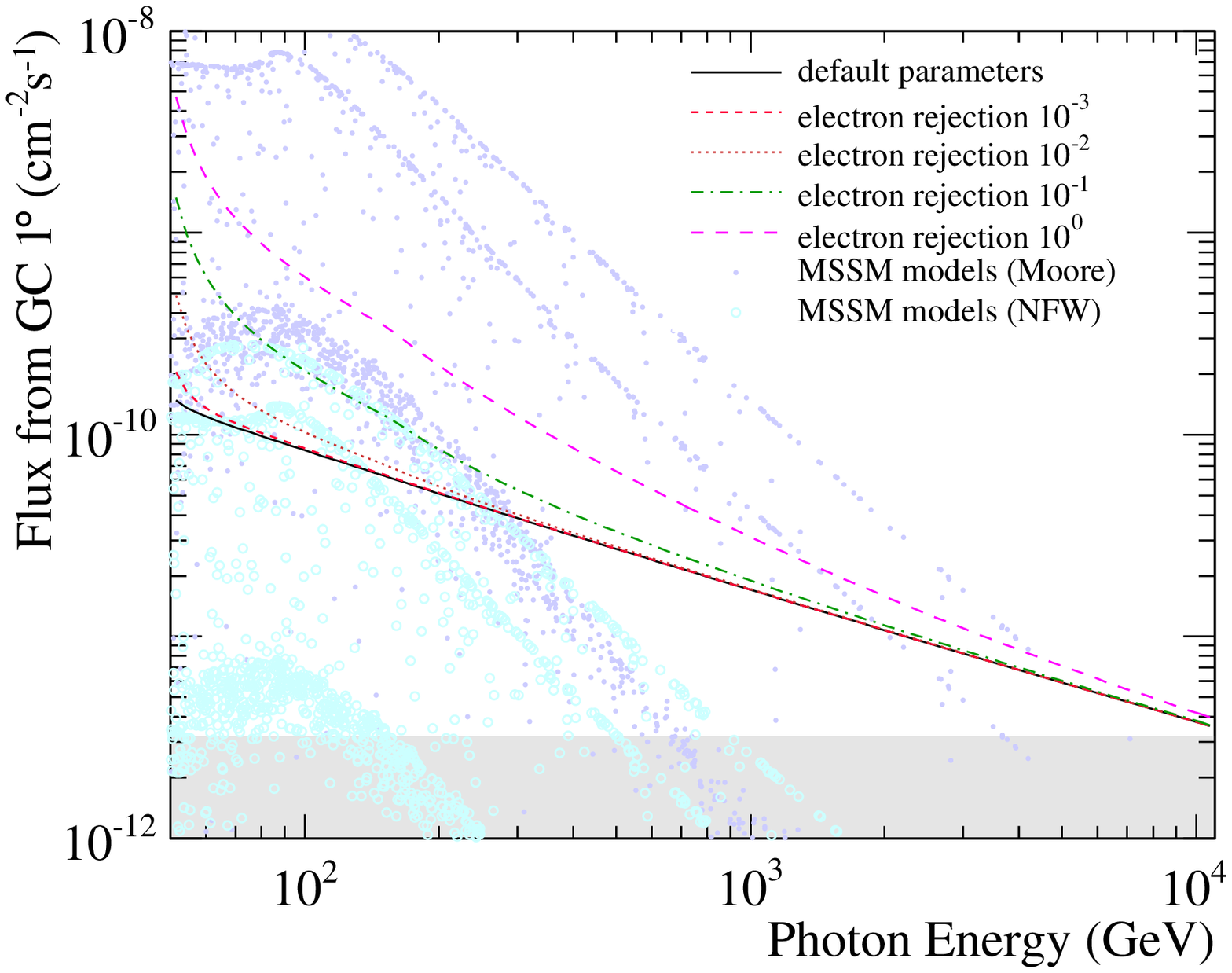} \includegraphics[%
  width=0.49\textwidth]{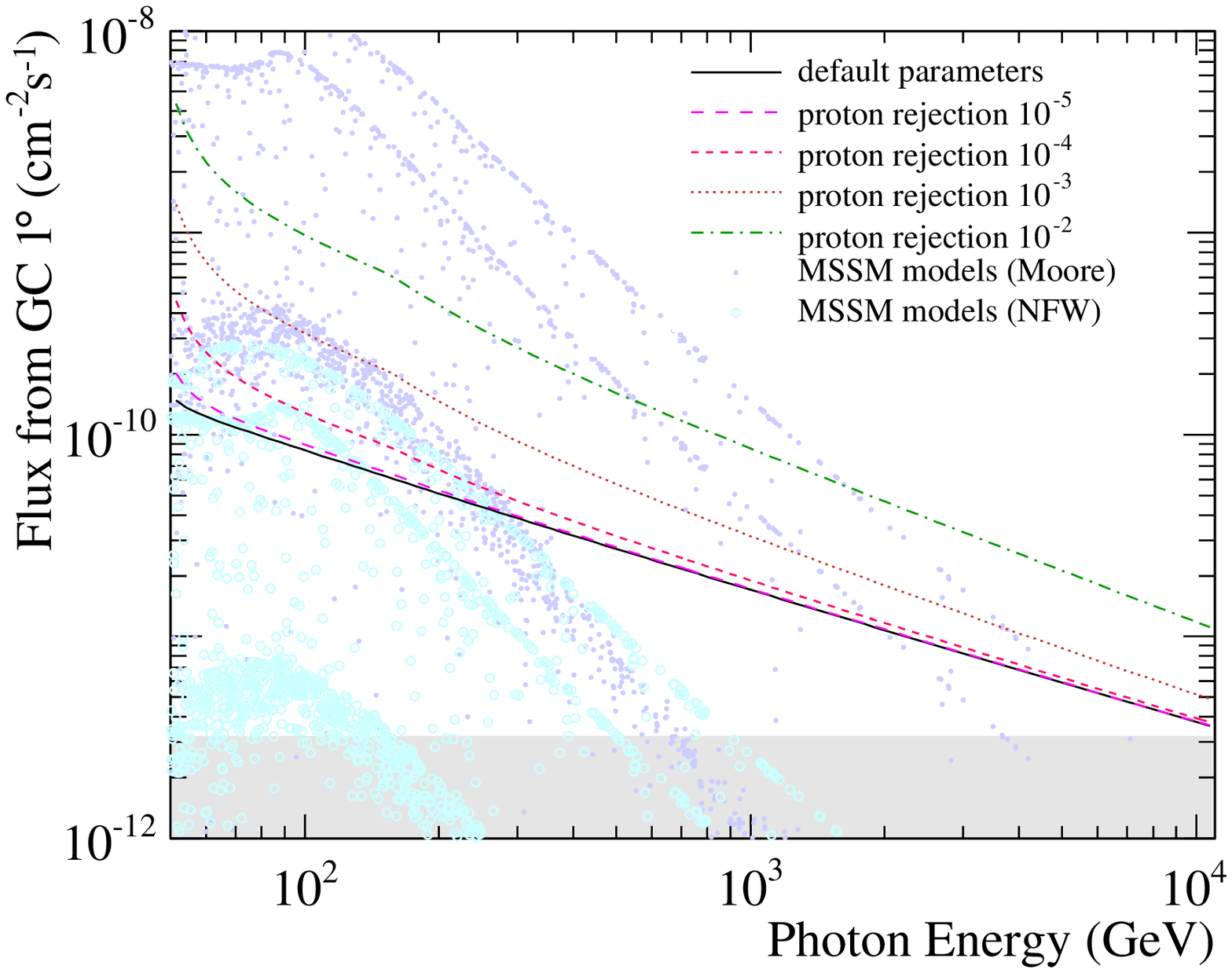}

\caption{\label{fig:ana-diffs}The sensitivity with different detector features:
energy resolution (top-left), angular resolution (top-right), electron
rejection (bottom-left), nuclei rejection (bottom-right). The default
settings are: energy resolution 1.5\%, angular resolution $0.5^{\circ}$,
electron rejection $10^{-4}$ and nuclei rejection $10^{-7}$. The
gray area at the bottom of each panel represents the flux corresponding
to less than one event per year (exposure time) for a 1 m$^{2}$ detector.}
\end{figure*}

The results of $f_{\textrm{min}}$ for different detector performance,
i.e., different energy resolution, angular resolution, electron rejection
and nuclei rejection respectively, are shown in Fig. \ref{fig:ana-diffs}.
The default detector performance settings are: energy resolution 1.5\%,
angular resolution $0.5^{\circ}$, electron rejection $10^{-4}$ and
nuclei rejection $10^{-7}$. The different performances used in the
calculation are summerized in \prettyref{tab:gc-parameters}. The
dots and triangles in the figure are the MSSM model predictions with
a random scan in the eight-dimensional parameter space as introduced
in Sec. \ref{sec:MSSM-dark-matter}. Both Moore and NFW profile are
calculated and plotted. It is shown that the default detector configuration
is powerful enough to probe much of the MSSM parameter space for the
model with DM density as cuspy as Moore type. If the DM density profile
is NFW-like the sensitivity will be worse. We will discuss the effects
of different density profiles more in the following.%
\begin{table}[hbt]

\caption{GC detectability calculation configuration\label{tab:gc-parameters}}

\begin{tabular}{r|>{\centering}p{0.50\columnwidth}}
\hline 
common setup&
\multicolumn{1}{p{0.50\columnwidth}}{\parbox[t]{0.50\columnwidth}{detector area: $\unit[1]{m^{2}}$, \\
exposure time: $\unit[7.9\times10^{6}]{s}$ ,\\
energy range: $\unit[50]{GeV}\sim\unit[10]{TeV}$}}\tabularnewline
\hline
energy resolution&
 1\%, 1.5\%, 2\%, 5\%, 10\%\tabularnewline
angular resolution&
 $0.1^{\circ}$, $0.3^{\circ}$, $0.5^{\circ}$, $1.0^{\circ}$, $3.0^{\circ}$\tabularnewline
electron rejection&
$10^{0}$, $10^{-1}$, $10^{-2}$, $10^{-3}$, $10^{-4}$\tabularnewline
proton rejection&
$10^{-2}$, $10^{-3}$, $10^{-4}$, $10^{-5}$, $10^{-7}$ \tabularnewline
\hline
\end{tabular}
\end{table}

\begin{figure}[hbt]
\includegraphics[%
  width=0.49\textwidth]{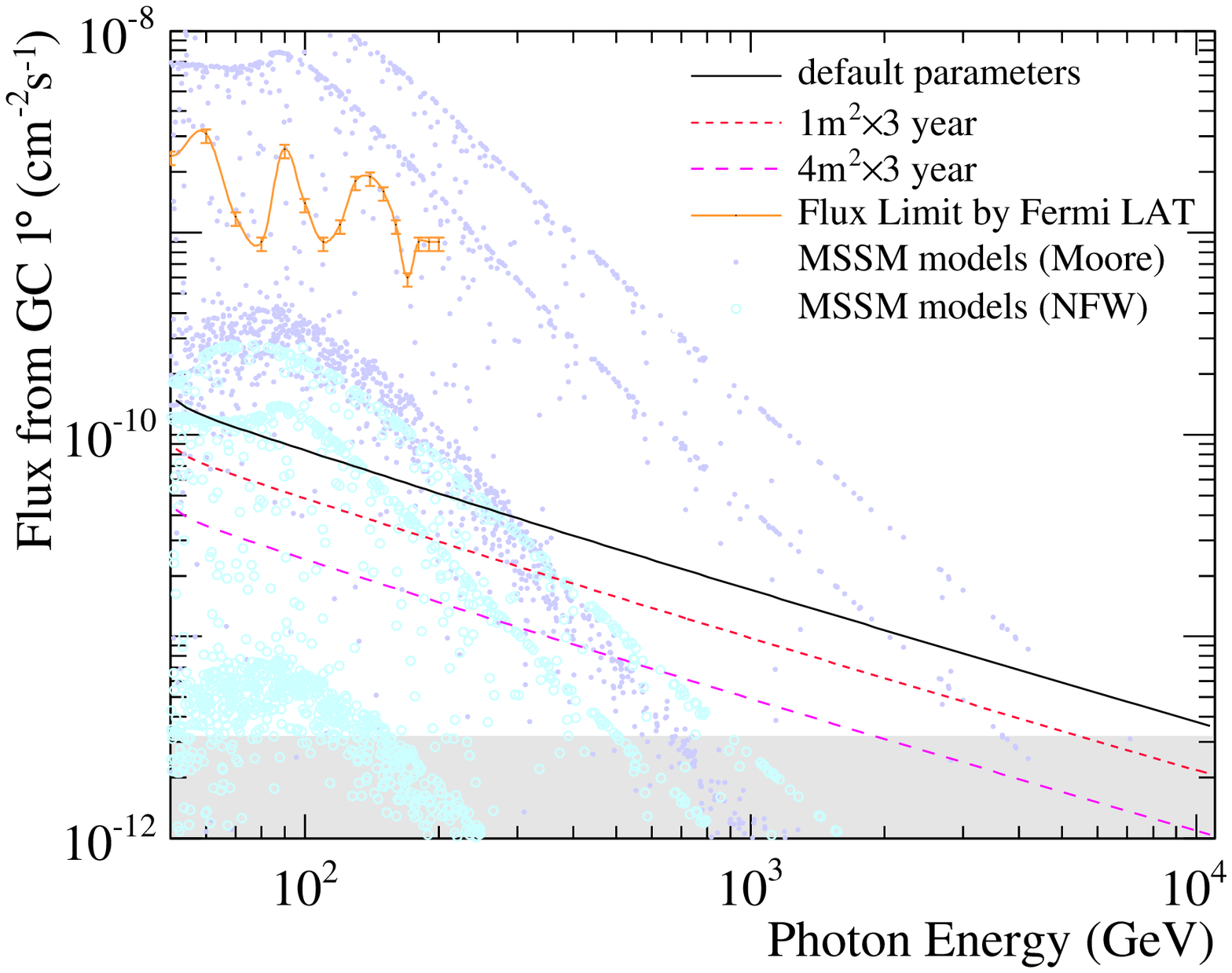}

\caption{The sensitivity with different exposure factor and detector area.\label{fig:bkg-D-real-type}}
\end{figure}

In Fig. \ref{fig:bkg-D-real-type} we show the comparison of sensitivities
for longer exposure and larger area of the same type detector. The
upper limits of monochromatic photons from $30$ to $200$ GeV derived
according to 11 month Fermi data are shown by the line with errorbars
\cite{2010PhRvL.104i1302A}. Note the observation region of Fermi's
result is $|b|>10^{\circ}$ plus $20^{\circ}\times20^{\circ}$ area
around the GC, which is much larger than 1 degree around GC adopted
here. Therefore the limits from GC would be smaller than the present
given values. We just plot the results here for reference.

\begin{figure*}[hbt]
\includegraphics[%
  width=0.49\textwidth,
  keepaspectratio]{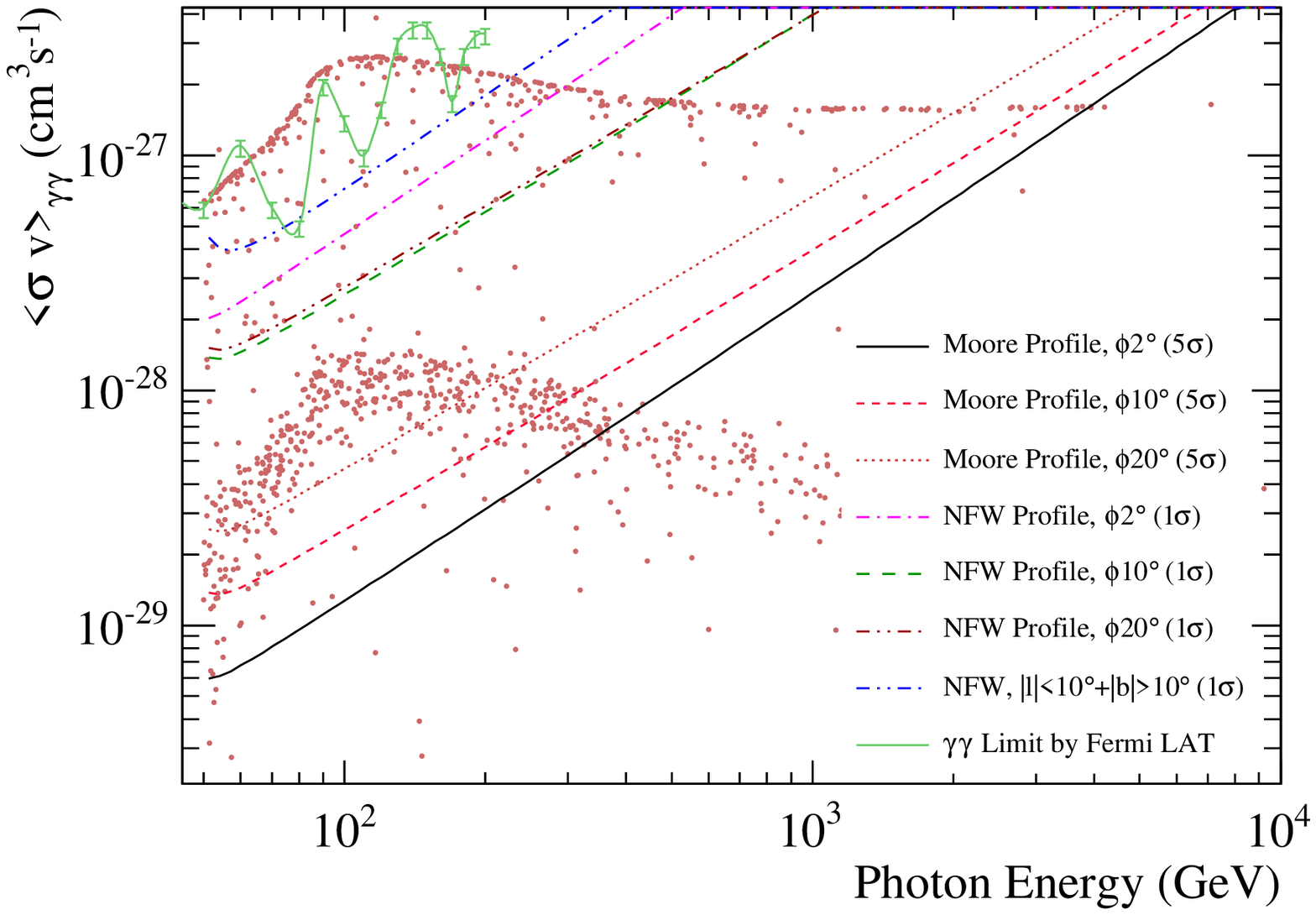}  \includegraphics[%
  width=0.49\textwidth,
  keepaspectratio]{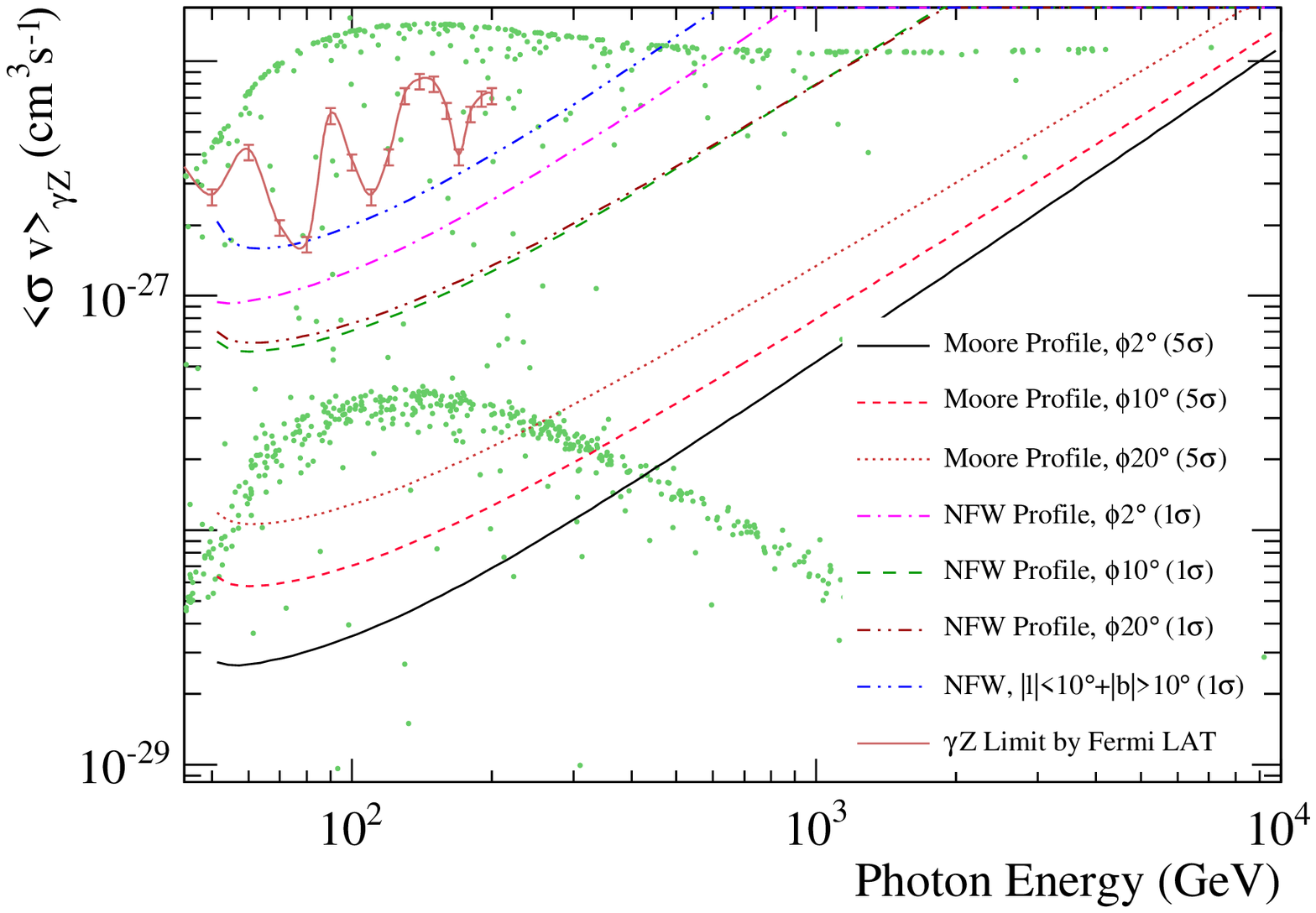}

\caption{The minimal detectable cross section for $\gamma\gamma$ (left) and
$\gamma Z_{0}$ (right) channels in different observation area \label{fig:cross}}
\end{figure*}

In Fig. \ref{fig:cross} the minimal detectable cross section is shown,
for $\gamma\gamma$ (left) and $\gamma Z_{0}$ (right) channels respectively.
In this plot the default detector performance is adopted, and we show
the results for the NFW and Moore profiles, with different sky regions.
It is shown that for the NFW profile the detectability will be orders
of magnitude worse than that for the Moore profile. We also note that
for the NFW profile the sensitivity is not affected significantly
by the choice of sky area if the GC is included. However, it is not
the case for the Moore profile. This can be understood that for the
Moore profile the signal flux drops very fast with the increase of
the angle to the GC, which leads to a fast decrease of the signal-to-noise
ratio. Therefore the large-angle average results in a worse sensitivity
than the case focusing on the small region around the GC. The $95\%$
exclusion limits of the cross section for the NFW profile from Fermi
are also plotted \cite{2010PhRvL.104i1302A}. To compare with the
capability of Fermi, we show the exclusion power of the detector with
the default parameters mentioned before, at $1\sigma$ level for the
same density profile and observational sky region as Fermi. 

\section{Whole sky scan sensitivity with Monte-Carlo simulation \label{sec:mc}}

In this section we discuss the all-sky observation with scan mode.
The effective exposure time is non-uniform at different directions
for specific orbit of the detector. Therefore the simple analytical
method discussed in the previous section does not work any more. We
turn to use fast Monte-Carlo method to simulate the counts of signal
and background, and calculate the sensitivity.

\subsection{Simulation configuration}

\label{sec:Simulation-configuration}
\begin{figure}[hbt]
\includegraphics[%
  width=0.49\textwidth]{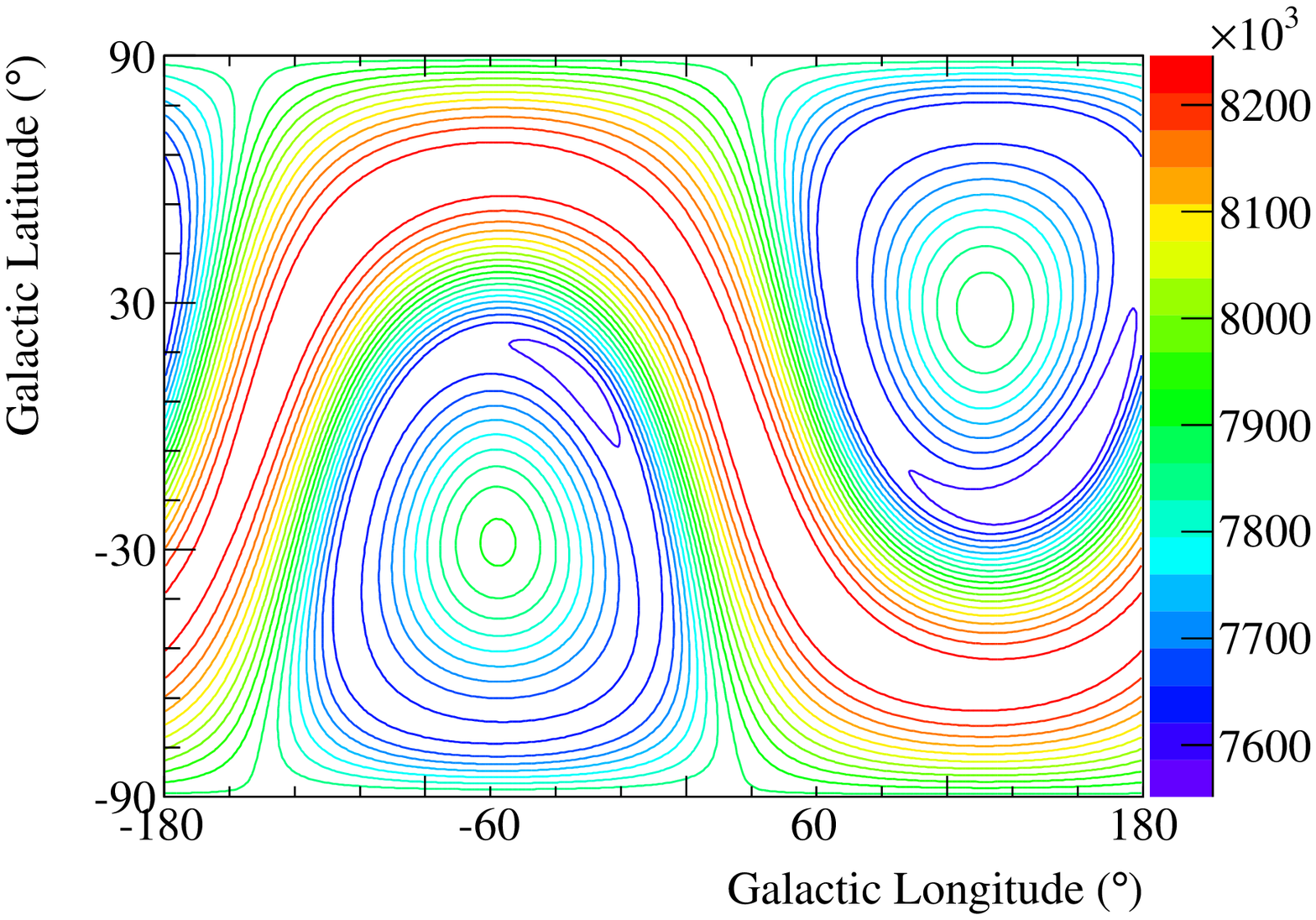}

\caption{\label{fig:Effective-survey-time}Sky-map of effective exposure time
according to the 2003 ISS orbit data. The whole sky area is divided
into $360\times180$ cells, with $1^{\circ}\times1^{\circ}$ of each
cell. This map is derived as follows. The effective exposure time
of each cell is calculated as $\cos(\theta)$ where $\theta$ is the
angle between the direction of the ISS and the direction of the cell.
If $\theta>\pi$ the effective exposure time is set to zero. Then
the effective time of each cell is calculated in an interval of one
second, and accumulated for all the time in one year ($3.15\times10^{7}{\textrm{s}}$).}
\end{figure}

The flight orbit is assumed to
be the orbit of ISS. We put an ideal detector on the orbit, and calculate
the signal and background particle counts by the Monte-Carlo sampling.
We take the orbit data of ISS in 2003 as an example. The orbit of
ISS has not changed too much in the period of several years. The sky-map
of exposure time is plotted in Fig. \ref{fig:Effective-survey-time}.

We choose detectors with a series of property parameters, including
energy resolution, field of view, electron and nuclei rejection powers.
Seven neutralino masses, between 50 GeV and 1200 GeV, are chosen in
the simulation. The details of the configurations are listed in Table
\ref{tab:Simulation-configuration}.

\begin{table}[hbt]

\caption{Survey simulation configuration\label{tab:Simulation-configuration}}

\begin{tabular}{r|>{\centering}p{0.50\columnwidth}}
\hline 
common configuration&
\multicolumn{1}{p{0.50\columnwidth}}{\parbox[t]{0.50\columnwidth}{geometry factor: 3.14 m$^{2}$sr, \\
flight time: 1 year, \\
flight orbit: ISS orbit}}\tabularnewline
\hline
simulated energy (GeV)&
\parbox[t]{0.50\columnwidth}{ 77.6, 199.7, 416.3, 611.4, \\
808.6, 1006.0, 1220.9}\tabularnewline
\hline
energy resolution&
 1\%, 1.5\%, 2\%\tabularnewline
field of view&
 $40^{\circ}$, $60^{\circ}$, $90^{\circ}$\tabularnewline
electron rejection&
 $10^{-2}$, $10^{-3}$, $10^{-4}$\tabularnewline
proton rejection&
 $10^{-5}$, $10^{-6}$, $10^{-7}$ \tabularnewline
\hline
\end{tabular}
\end{table}

To get better statistics, a large enough cross section was chosen
in the simulation. The photons from $\gamma\gamma$ and $\gamma Z_{0}$
annihilation channels are not distinguished, and only a sensitivity
for single gamma line emission was given. Actually the photons from
these two channels can not be discriminated when neutralino mass is
larger than several hundred GeV, and if neutralino mass is small,
there could be two peeks on the photon energy spectrum, thus the sensitivity
would be higher.

The overall contribution from DM annihilation, not only the smooth
halo, but also the substructures, is taken into account. In the simulation
we use Moore profile for both the smooth halo and subhalos. To compare
with theory predictions, the model points are plotted along with the
minimum detectable curves. In the calculation both Moore and NFW profile
are used. And also the contribution of substructures and smooth halo
are summed together.

\subsection{Event generation and reconstruction simulation}

The basic procedure of the simulation can be divided into two phases.
First is the generation of the event samples according to the angular
and energy distributions of each component, including the signal and
backgrounds. The second phase is to simulate the event reconstruct
process based on the detector performances.

For the signal events, the energy of generated photon is monochromatic
for a given neutralino mass. The spatial distribution follows the
result as shown in Fig. \ref{fig:dm-dist}. The energy spectra and
angular distributions of backgrounds are described in Sec. \ref{sec:Backgrounds}.
In the simulation, we also need to know the direction of the detector.
The detector pointing direction is set to be the same as the direction
of ISS.

The events are generated in interval of one second in detector operation
time. To simplify the simulation code, static event rate for each
component is used. The event rate is set to be the maximum value that
can be received by the detector with acceptance $4\pi\cdot A$, for
energy range from 30 GeV to 1400 GeV, and a Poisson smear is used
when event number in the one-second-interval is too small. However,
the actual events received by a specific detector will be much smaller
than the above value since the effective area for photons with large
incident angle will decrease. This effect is taken into account at
the second phase through an efficiency factor $\eta$, defined as
$\eta_{i}\cos(p)$ when $p<p_{FOV}$, and $0$ when $p\geq p_{FOV}$,
where $p$ is the angle between the incident photon and the detector
pointing direction, $p_{FOV}$ is the maximum receiving angle of incident
particles, and $\eta_{i}$ is the selection efficiency for different
incident particle. For photons the selection efficiency usually assumed
to be 90\%. For electrons and nuclei events, the detector rejection
power is used as the selection efficiency. Then one event is kept
with probability $\eta$.

\begin{figure}[hbt]
\includegraphics[%
  width=0.49\textwidth]{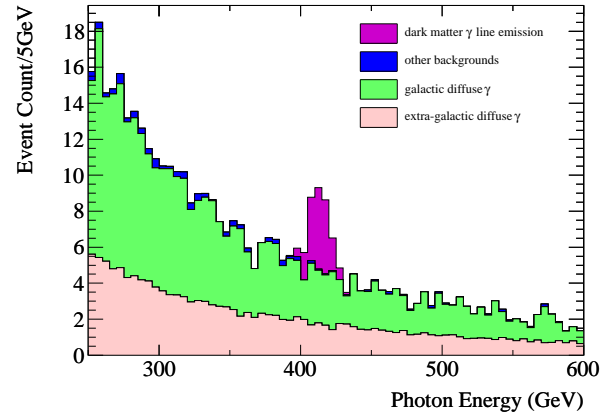}

\caption{A sample reconstructed stacked photon energy spectrum, for neutralino
mass 416 GeV.}

\label{fig:sample}
\end{figure}

The events kept in the efficiency selection are then reconstructed.
The reconstruction of energy is simply implemented by a Gaussian smear
of the inject photon energy. The Gaussian width is $\sigma_{e}$,
which represents the energy resolution of the detector. The nuclei
rejection and electron rejection are simulated as a fixed pass rate
as mentioned before. Passed electrons and nuclei are reconstructed
the same way as photons. \prettyref{fig:sample} shows a sample reconstructed
energy spectrum for the neutralino mass 416 GeV. The direction reconstruction
simulation is in the same way as energy reconstruction, with two-dimensional
Gaussian smear. 

\subsection{Results of simulation }

\begin{figure*}[hbt]
\includegraphics[%
  width=0.49\textwidth]{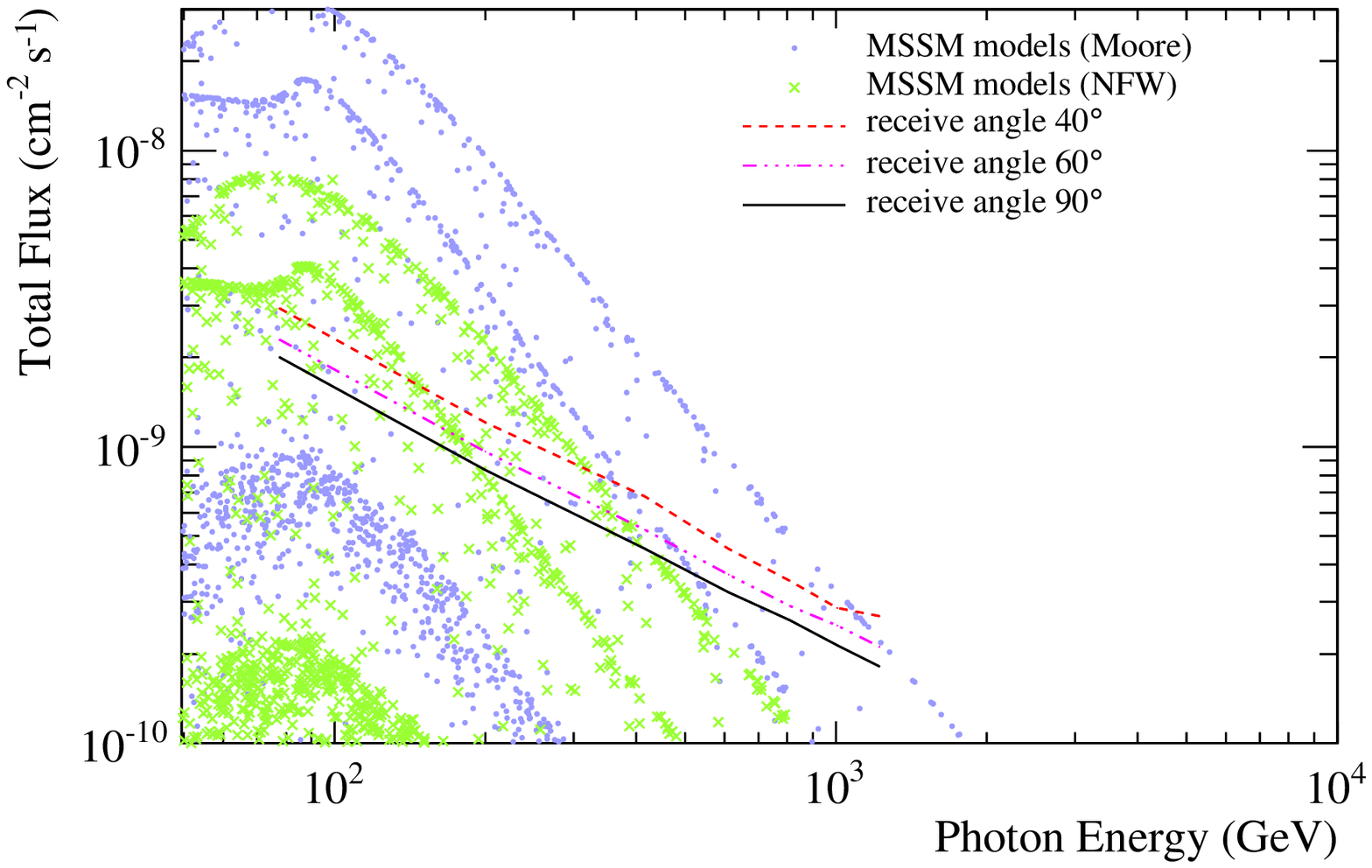} \includegraphics[%
  width=0.49\textwidth]{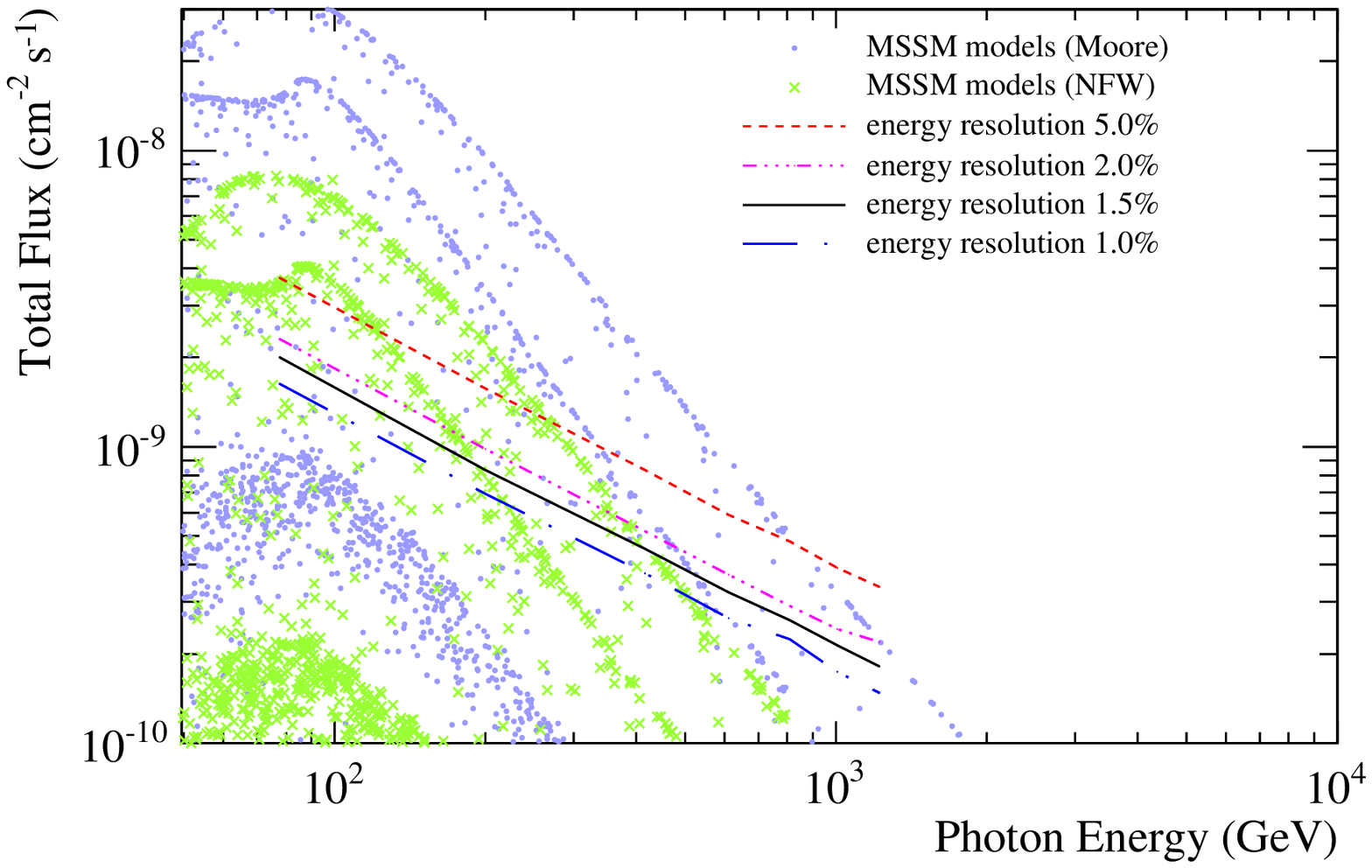} \includegraphics[%
  width=0.49\textwidth]{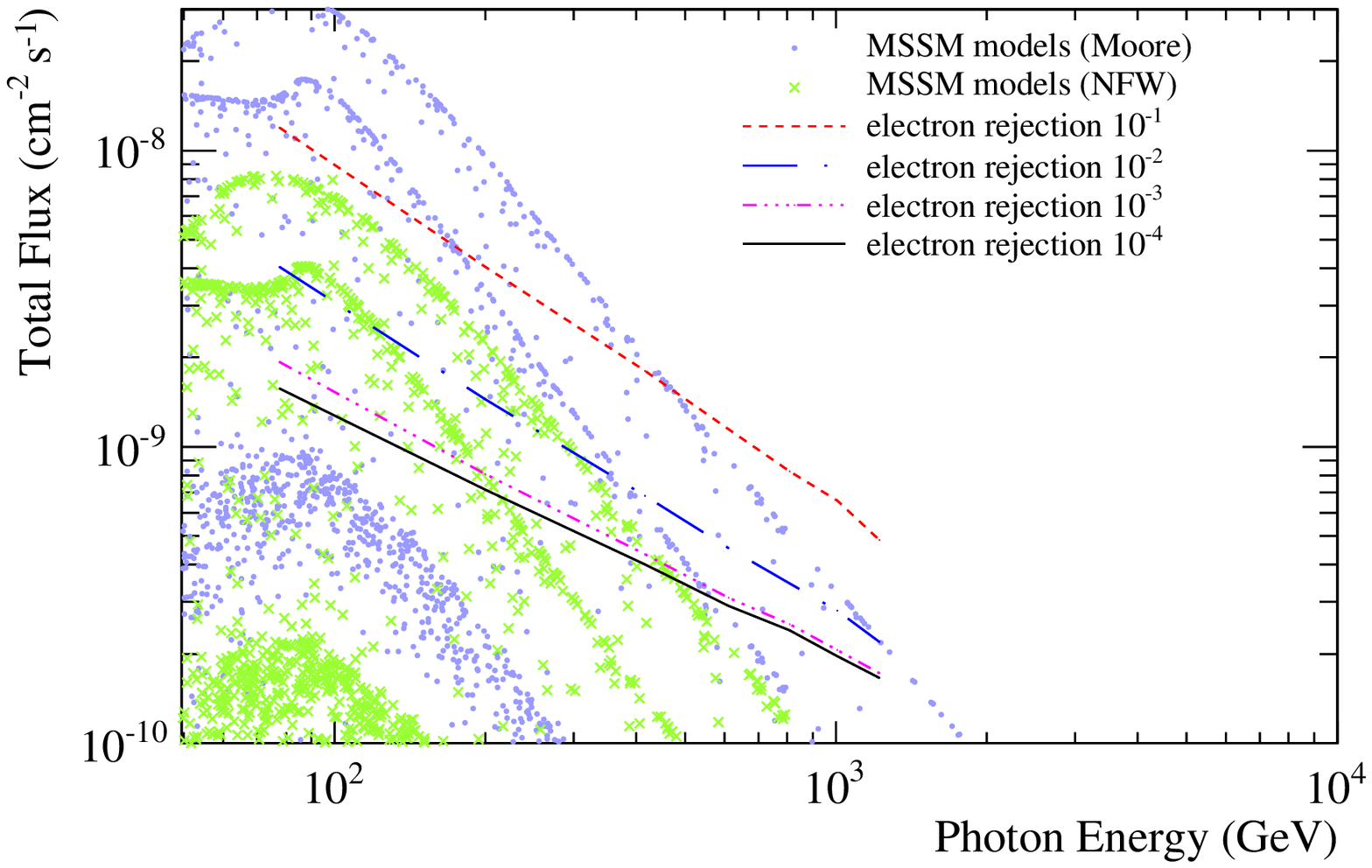} \includegraphics[%
  width=0.49\textwidth]{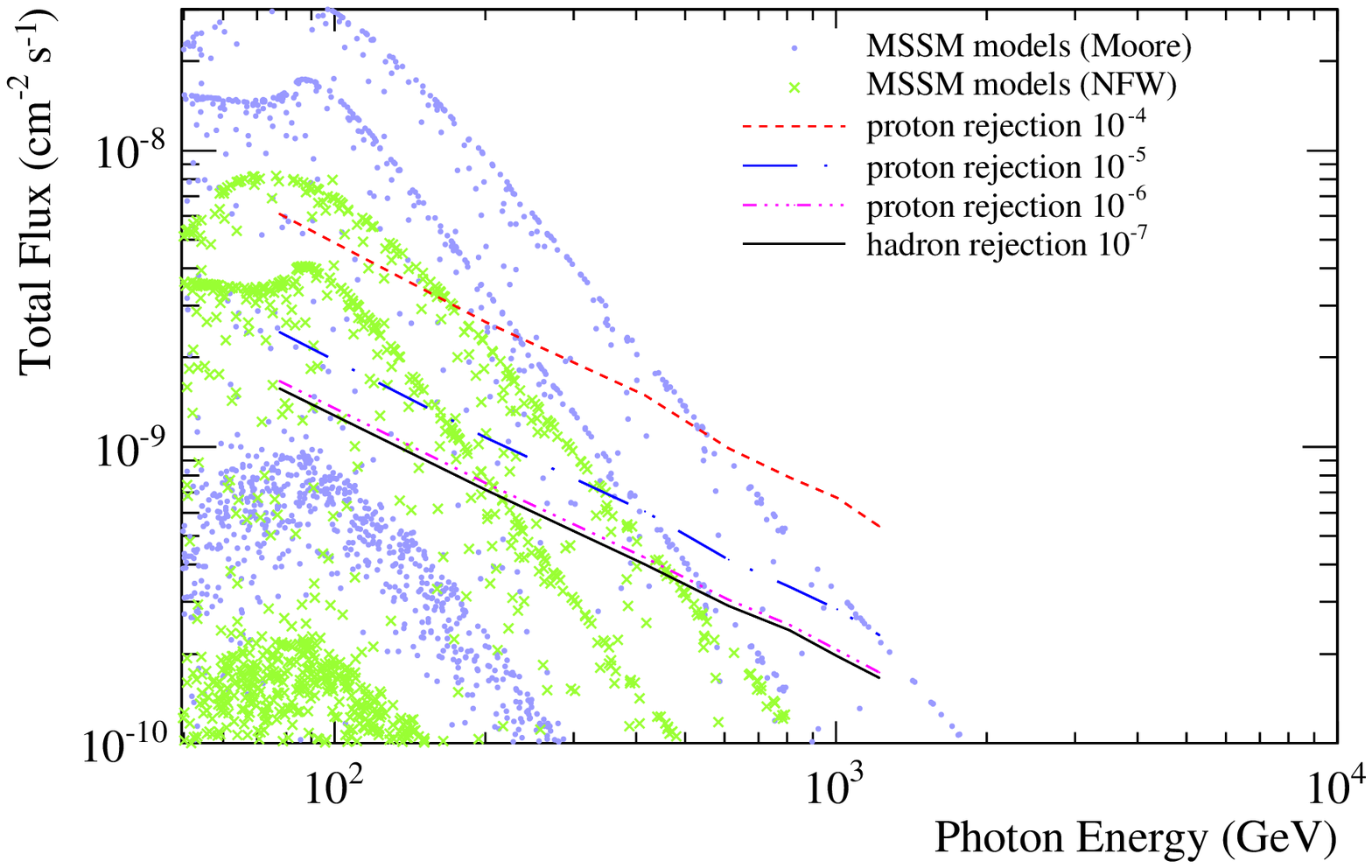}

\caption{\label{fig:scan}The minimum detectable flux of DM line emission
in scan mode. Top-left: comparison among different maximum receiving
angle (field of view); top-right: comparison among different energy
resolution; bottom-left: comparison among different electron rejection
power; bottom-left: comparison among different nuclei rejection power.
The default parameters are energy resolution 1.5\%, maximum receiving
angle $90^{\circ}$, electron rejection $10^{-4}$, and proton rejection
$10^{-7}$.}
\end{figure*}

The results of sensitivities for scan observation are shown in Fig.
\ref{fig:scan}, for different detector performance. The area of detector
is adopted to be 1 m $\times$ 1 m. The points represent the MSSM
model predicted fluxes of the all-sky DM annihilation, with substructures
included, for both Moore and NFW profile. Compared with the results
of GC, the scan mode will be less sensitive for detecting the DM annihilation
signal by more than one order of magnitude.

\section{Discussion and conclusion\label{sec:discuss}}

In this work we studied the strategy and feasibility of the monochromatic
$\gamma$-ray detection from DM annihilation. We studied the detectability
of the monochromatic $\gamma$-rays from the GC and from the dark
matter halo and subhalos, by analytical analysis and numerical Monte-Carlo
simulation. The detector performance is studied for the purpose of
monochromatic $\gamma$-ray detection. 

The sensitivity depends on the observation region and the DM density
profiles. According to the Moore profile, the sensitivity of GC region
would be much higher than other region. However, this is not true
for NFW-like profiles. As we currently have little knowledge about
the actual distribution, we can not rely on a particular model. Until
now, high energy gamma observations, such HESS and Fermi, have not
seen any evidence in the GC region. A safe strategy could be starting
from the scan mode, then focusing on the interesting area if any evidence
is found. On the other hand, the GC is always an interesting region
for possible DM signals. The observation of the GC should be tuned
as much as possible.

In order to detect the monochromatic $\gamma$-ray emission of DM
annihilation in scan mode, the energy resolution and geometry factor,
both detector area and field of view, are the most important detector
feature. However, The resources are usually very limited on orbit,
on both detector mass and power supply. With a given mass of the detector,
thinner calorimeter would lead to larger active area. On the other
hand, thinner calorimeters usually have poorer energy resolution.
Therefore, to get better sensitivity, a careful balance is needed
in detector designing. Also any detector design that would limit the
field of view of the detector should better be avoided. At low energies
($<300$GeV), the ability to reject cosmic-ray background, electrons
and nuclei, are also important. To keep cosmic-ray background low
enough, that is, lower than the major background, diffuse $\gamma$,
at least $10^{-3}$ electron rejection and $10^{-6}$ proton rejection
is required.

If we are focusing on GC or other interesting region in pointing mode,
the angular resolution, instead of field of view, becomes very important.
In the GC region, the $\gamma$-ray emission is complex: several point
sources exist there. Thus good angular resolution is needed to separate
them apart, and suppress these point source backgrounds. Of course
the energy resolution and detector area are still important features
in order to detector monochromatic $\gamma$-rays. Also, at low energy,
the rejection power requirements of electron and proton are the same
as in scan mode.

\begin{acknowledgments}
This work is supported by Natural Sciences Foundation of China (No.
10435070, No. 10773011, No. 10721140381, No. 10099630) and by China
Ministry of Science and Technology (No. 2007CB16101, No. 2010CB833000).
\end{acknowledgments}
\bibliographystyle{apsrev}

\begin{thebibliography}{34}
\expandafter\ifx\csname natexlab\endcsname\relax\def\natexlab#1{#1}\fi
\expandafter\ifx\csname bibnamefont\endcsname\relax
  \def\bibnamefont#1{#1}\fi
\expandafter\ifx\csname bibfnamefont\endcsname\relax
  \def\bibfnamefont#1{#1}\fi
\expandafter\ifx\csname citenamefont\endcsname\relax
  \def\citenamefont#1{#1}\fi
\expandafter\ifx\csname url\endcsname\relax
  \def\url#1{\texttt{#1}}\fi
\expandafter\ifx\csname urlprefix\endcsname\relax\def\urlprefix{URL }\fi
\providecommand{\bibinfo}[2]{#2}
\providecommand{\eprint}[2][]{\url{#2}}

\bibitem[{\citenamefont{{Gondolo} et~al.}(2004)\citenamefont{{Gondolo},
  {Edsj{\"o}}, {Ullio}, {Bergstr{\"o}m}, {Schelke}, and
  {Baltz}}}]{2004JCAP...07..008G}
\bibinfo{author}{\bibfnamefont{P.}~\bibnamefont{{Gondolo}}},
  \bibinfo{author}{\bibfnamefont{J.}~\bibnamefont{{Edsj{\"o}}}},
  \bibinfo{author}{\bibfnamefont{P.}~\bibnamefont{{Ullio}}},
  \bibinfo{author}{\bibfnamefont{L.}~\bibnamefont{{Bergstr{\"o}m}}},
  \bibinfo{author}{\bibfnamefont{M.}~\bibnamefont{{Schelke}}},
  \bibnamefont{and} \bibinfo{author}{\bibfnamefont{E.~A.}
  \bibnamefont{{Baltz}}}, \bibinfo{journal}{Journal of Cosmology and
  Astro-Particle Physics} \textbf{\bibinfo{volume}{7}}, \bibinfo{pages}{8}
  (\bibinfo{year}{2004}), \eprint{arXiv:astro-ph/0406204}.

\bibitem[{\citenamefont{{Catena} and {Ullio}}(2009)}]{2009arXiv0907.0018C}
\bibinfo{author}{\bibfnamefont{R.}~\bibnamefont{{Catena}}} \bibnamefont{and}
  \bibinfo{author}{\bibfnamefont{P.}~\bibnamefont{{Ullio}}},
  \bibinfo{journal}{ArXiv e-prints}  (\bibinfo{year}{2009}),
  \eprint{0907.0018}.

\bibitem[{\citenamefont{{Salucci} et~al.}(2010)\citenamefont{{Salucci},
  {Nesti}, {Gentile}, and {Martins}}}]{2010arXiv1003.3101S}
\bibinfo{author}{\bibfnamefont{P.}~\bibnamefont{{Salucci}}},
  \bibinfo{author}{\bibfnamefont{F.}~\bibnamefont{{Nesti}}},
  \bibinfo{author}{\bibfnamefont{G.}~\bibnamefont{{Gentile}}},
  \bibnamefont{and} \bibinfo{author}{\bibfnamefont{C.~F.}
  \bibnamefont{{Martins}}}, \bibinfo{journal}{ArXiv e-prints}
  (\bibinfo{year}{2010}), \eprint{1003.3101}.

\bibitem[{\citenamefont{{Komatsu} et~al.}(2010)\citenamefont{{Komatsu},
  {Smith}, {Dunkley}, {Bennett}, {Gold}, {Hinshaw}, {Jarosik}, {Larson},
  {Nolta}, {Page} et~al.}}]{2010arXiv1001.4538K}
\bibinfo{author}{\bibfnamefont{E.}~\bibnamefont{{Komatsu}}},
  \bibinfo{author}{\bibfnamefont{K.~M.} \bibnamefont{{Smith}}},
  \bibinfo{author}{\bibfnamefont{J.}~\bibnamefont{{Dunkley}}},
  \bibinfo{author}{\bibfnamefont{C.~L.} \bibnamefont{{Bennett}}},
  \bibinfo{author}{\bibfnamefont{B.}~\bibnamefont{{Gold}}},
  \bibinfo{author}{\bibfnamefont{G.}~\bibnamefont{{Hinshaw}}},
  \bibinfo{author}{\bibfnamefont{N.}~\bibnamefont{{Jarosik}}},
  \bibinfo{author}{\bibfnamefont{D.}~\bibnamefont{{Larson}}},
  \bibinfo{author}{\bibfnamefont{M.~R.} \bibnamefont{{Nolta}}},
  \bibinfo{author}{\bibfnamefont{L.}~\bibnamefont{{Page}}},
  \bibnamefont{et~al.}, \bibinfo{journal}{ArXiv e-prints}
  (\bibinfo{year}{2010}), \eprint{1001.4538}.

\bibitem[{\citenamefont{{Navarro} et~al.}(1997)\citenamefont{{Navarro},
  {Frenk}, and {White}}}]{1997ApJ...490..493N}
\bibinfo{author}{\bibfnamefont{J.~F.} \bibnamefont{{Navarro}}},
  \bibinfo{author}{\bibfnamefont{C.~S.} \bibnamefont{{Frenk}}},
  \bibnamefont{and} \bibinfo{author}{\bibfnamefont{S.~D.~M.}
  \bibnamefont{{White}}}, \bibinfo{journal}{\apj}
  \textbf{\bibinfo{volume}{490}}, \bibinfo{pages}{493} (\bibinfo{year}{1997}),
  \eprint{arXiv:astro-ph/9611107}.

\bibitem[{\citenamefont{{Moore}
  et~al.}(1999{\natexlab{a}})\citenamefont{{Moore}, {Quinn}, {Governato},
  {Stadel}, and {Lake}}}]{1999MNRAS.310.1147M}
\bibinfo{author}{\bibfnamefont{B.}~\bibnamefont{{Moore}}},
  \bibinfo{author}{\bibfnamefont{T.}~\bibnamefont{{Quinn}}},
  \bibinfo{author}{\bibfnamefont{F.}~\bibnamefont{{Governato}}},
  \bibinfo{author}{\bibfnamefont{J.}~\bibnamefont{{Stadel}}}, \bibnamefont{and}
  \bibinfo{author}{\bibfnamefont{G.}~\bibnamefont{{Lake}}},
  \bibinfo{journal}{\mnras} \textbf{\bibinfo{volume}{310}},
  \bibinfo{pages}{1147} (\bibinfo{year}{1999}{\natexlab{a}}),
  \eprint{arXiv:astro-ph/9903164}.

\bibitem[{\citenamefont{{Navarro} et~al.}(2010)\citenamefont{{Navarro},
  {Ludlow}, {Springel}, {Wang}, {Vogelsberger}, {White}, {Jenkins}, {Frenk},
  and {Helmi}}}]{2010MNRAS.402...21N}
\bibinfo{author}{\bibfnamefont{J.~F.} \bibnamefont{{Navarro}}},
  \bibinfo{author}{\bibfnamefont{A.}~\bibnamefont{{Ludlow}}},
  \bibinfo{author}{\bibfnamefont{V.}~\bibnamefont{{Springel}}},
  \bibinfo{author}{\bibfnamefont{J.}~\bibnamefont{{Wang}}},
  \bibinfo{author}{\bibfnamefont{M.}~\bibnamefont{{Vogelsberger}}},
  \bibinfo{author}{\bibfnamefont{S.~D.~M.} \bibnamefont{{White}}},
  \bibinfo{author}{\bibfnamefont{A.}~\bibnamefont{{Jenkins}}},
  \bibinfo{author}{\bibfnamefont{C.~S.} \bibnamefont{{Frenk}}},
  \bibnamefont{and} \bibinfo{author}{\bibfnamefont{A.}~\bibnamefont{{Helmi}}},
  \bibinfo{journal}{\mnras} \textbf{\bibinfo{volume}{402}}, \bibinfo{pages}{21}
  (\bibinfo{year}{2010}), \eprint{0810.1522}.

\bibitem[{\citenamefont{{Jing} and {Suto}}(2000)}]{2000ApJ...529L..69J}
\bibinfo{author}{\bibfnamefont{Y.~P.} \bibnamefont{{Jing}}} \bibnamefont{and}
  \bibinfo{author}{\bibfnamefont{Y.}~\bibnamefont{{Suto}}},
  \bibinfo{journal}{\apjl} \textbf{\bibinfo{volume}{529}}, \bibinfo{pages}{L69}
  (\bibinfo{year}{2000}), \eprint{arXiv:astro-ph/9909478}.

\bibitem[{\citenamefont{{Berezinsky} et~al.}(1992)\citenamefont{{Berezinsky},
  {Gurevich}, and {Zybin}}}]{1992PhLB..294..221B}
\bibinfo{author}{\bibfnamefont{V.~S.} \bibnamefont{{Berezinsky}}},
  \bibinfo{author}{\bibfnamefont{A.~V.} \bibnamefont{{Gurevich}}},
  \bibnamefont{and} \bibinfo{author}{\bibfnamefont{K.~P.}
  \bibnamefont{{Zybin}}}, \bibinfo{journal}{Physics Letters B}
  \textbf{\bibinfo{volume}{294}}, \bibinfo{pages}{221} (\bibinfo{year}{1992}).

\bibitem[{\citenamefont{{Lavalle} et~al.}(2008)\citenamefont{{Lavalle}, {Yuan},
  {Maurin}, and {Bi}}}]{2008A&A...479..427L}
\bibinfo{author}{\bibfnamefont{J.}~\bibnamefont{{Lavalle}}},
  \bibinfo{author}{\bibfnamefont{Q.}~\bibnamefont{{Yuan}}},
  \bibinfo{author}{\bibfnamefont{D.}~\bibnamefont{{Maurin}}}, \bibnamefont{and}
  \bibinfo{author}{\bibfnamefont{X.}~\bibnamefont{{Bi}}},
  \bibinfo{journal}{\aap} \textbf{\bibinfo{volume}{479}}, \bibinfo{pages}{427}
  (\bibinfo{year}{2008}), \eprint{0709.3634}.

\bibitem[{\citenamefont{{Tormen} et~al.}(1998)\citenamefont{{Tormen},
  {Diaferio}, and {Syer}}}]{1998MNRAS.299..728T}
\bibinfo{author}{\bibfnamefont{G.}~\bibnamefont{{Tormen}}},
  \bibinfo{author}{\bibfnamefont{A.}~\bibnamefont{{Diaferio}}},
  \bibnamefont{and} \bibinfo{author}{\bibfnamefont{D.}~\bibnamefont{{Syer}}},
  \bibinfo{journal}{\mnras} \textbf{\bibinfo{volume}{299}},
  \bibinfo{pages}{728} (\bibinfo{year}{1998}), \eprint{arXiv:astro-ph/9712222}.

\bibitem[{\citenamefont{{Moore}
  et~al.}(1999{\natexlab{b}})\citenamefont{{Moore}, {Ghigna}, {Governato},
  {Lake}, {Quinn}, {Stadel}, and {Tozzi}}}]{1999ApJ...524L..19M}
\bibinfo{author}{\bibfnamefont{B.}~\bibnamefont{{Moore}}},
  \bibinfo{author}{\bibfnamefont{S.}~\bibnamefont{{Ghigna}}},
  \bibinfo{author}{\bibfnamefont{F.}~\bibnamefont{{Governato}}},
  \bibinfo{author}{\bibfnamefont{G.}~\bibnamefont{{Lake}}},
  \bibinfo{author}{\bibfnamefont{T.}~\bibnamefont{{Quinn}}},
  \bibinfo{author}{\bibfnamefont{J.}~\bibnamefont{{Stadel}}}, \bibnamefont{and}
  \bibinfo{author}{\bibfnamefont{P.}~\bibnamefont{{Tozzi}}},
  \bibinfo{journal}{\apjl} \textbf{\bibinfo{volume}{524}}, \bibinfo{pages}{L19}
  (\bibinfo{year}{1999}{\natexlab{b}}), \eprint{arXiv:astro-ph/9907411}.

\bibitem[{\citenamefont{{Ghigna} et~al.}(2000)\citenamefont{{Ghigna}, {Moore},
  {Governato}, {Lake}, {Quinn}, and {Stadel}}}]{2000ApJ...544..616G}
\bibinfo{author}{\bibfnamefont{S.}~\bibnamefont{{Ghigna}}},
  \bibinfo{author}{\bibfnamefont{B.}~\bibnamefont{{Moore}}},
  \bibinfo{author}{\bibfnamefont{F.}~\bibnamefont{{Governato}}},
  \bibinfo{author}{\bibfnamefont{G.}~\bibnamefont{{Lake}}},
  \bibinfo{author}{\bibfnamefont{T.}~\bibnamefont{{Quinn}}}, \bibnamefont{and}
  \bibinfo{author}{\bibfnamefont{J.}~\bibnamefont{{Stadel}}},
  \bibinfo{journal}{\apj} \textbf{\bibinfo{volume}{544}}, \bibinfo{pages}{616}
  (\bibinfo{year}{2000}), \eprint{arXiv:astro-ph/9910166}.

\bibitem[{\citenamefont{{Zentner} and {Bullock}}(2003)}]{2003ApJ...598...49Z}
\bibinfo{author}{\bibfnamefont{A.~R.} \bibnamefont{{Zentner}}}
  \bibnamefont{and} \bibinfo{author}{\bibfnamefont{J.~S.}
  \bibnamefont{{Bullock}}}, \bibinfo{journal}{\apj}
  \textbf{\bibinfo{volume}{598}}, \bibinfo{pages}{49} (\bibinfo{year}{2003}),
  \eprint{arXiv:astro-ph/0304292}.

\bibitem[{\citenamefont{{Gao} et~al.}(2004)\citenamefont{{Gao}, {White},
  {Jenkins}, {Stoehr}, and {Springel}}}]{2004MNRAS.355..819G}
\bibinfo{author}{\bibfnamefont{L.}~\bibnamefont{{Gao}}},
  \bibinfo{author}{\bibfnamefont{S.~D.~M.} \bibnamefont{{White}}},
  \bibinfo{author}{\bibfnamefont{A.}~\bibnamefont{{Jenkins}}},
  \bibinfo{author}{\bibfnamefont{F.}~\bibnamefont{{Stoehr}}}, \bibnamefont{and}
  \bibinfo{author}{\bibfnamefont{V.}~\bibnamefont{{Springel}}},
  \bibinfo{journal}{\mnras} \textbf{\bibinfo{volume}{355}},
  \bibinfo{pages}{819} (\bibinfo{year}{2004}), \eprint{arXiv:astro-ph/0404589}.

\bibitem[{\citenamefont{{Diemand} et~al.}(2004)\citenamefont{{Diemand},
  {Moore}, and {Stadel}}}]{2004MNRAS.352..535D}
\bibinfo{author}{\bibfnamefont{J.}~\bibnamefont{{Diemand}}},
  \bibinfo{author}{\bibfnamefont{B.}~\bibnamefont{{Moore}}}, \bibnamefont{and}
  \bibinfo{author}{\bibfnamefont{J.}~\bibnamefont{{Stadel}}},
  \bibinfo{journal}{\mnras} \textbf{\bibinfo{volume}{352}},
  \bibinfo{pages}{535} (\bibinfo{year}{2004}), \eprint{arXiv:astro-ph/0402160}.

\bibitem[{\citenamefont{{Springel} et~al.}(2008)\citenamefont{{Springel},
  {Wang}, {Vogelsberger}, {Ludlow}, {Jenkins}, {Helmi}, {Navarro}, {Frenk}, and
  {White}}}]{2008MNRAS.391.1685S}
\bibinfo{author}{\bibfnamefont{V.}~\bibnamefont{{Springel}}},
  \bibinfo{author}{\bibfnamefont{J.}~\bibnamefont{{Wang}}},
  \bibinfo{author}{\bibfnamefont{M.}~\bibnamefont{{Vogelsberger}}},
  \bibinfo{author}{\bibfnamefont{A.}~\bibnamefont{{Ludlow}}},
  \bibinfo{author}{\bibfnamefont{A.}~\bibnamefont{{Jenkins}}},
  \bibinfo{author}{\bibfnamefont{A.}~\bibnamefont{{Helmi}}},
  \bibinfo{author}{\bibfnamefont{J.~F.} \bibnamefont{{Navarro}}},
  \bibinfo{author}{\bibfnamefont{C.~S.} \bibnamefont{{Frenk}}},
  \bibnamefont{and} \bibinfo{author}{\bibfnamefont{S.~D.~M.}
  \bibnamefont{{White}}}, \bibinfo{journal}{\mnras}
  \textbf{\bibinfo{volume}{391}}, \bibinfo{pages}{1685} (\bibinfo{year}{2008}),
  \eprint{0809.0898}.

\bibitem[{\citenamefont{{Bullock} et~al.}(2001)\citenamefont{{Bullock},
  {Kolatt}, {Sigad}, {Somerville}, {Kravtsov}, {Klypin}, {Primack}, and
  {Dekel}}}]{2001MNRAS.321..559B}
\bibinfo{author}{\bibfnamefont{J.~S.} \bibnamefont{{Bullock}}},
  \bibinfo{author}{\bibfnamefont{T.~S.} \bibnamefont{{Kolatt}}},
  \bibinfo{author}{\bibfnamefont{Y.}~\bibnamefont{{Sigad}}},
  \bibinfo{author}{\bibfnamefont{R.~S.} \bibnamefont{{Somerville}}},
  \bibinfo{author}{\bibfnamefont{A.~V.} \bibnamefont{{Kravtsov}}},
  \bibinfo{author}{\bibfnamefont{A.~A.} \bibnamefont{{Klypin}}},
  \bibinfo{author}{\bibfnamefont{J.~R.} \bibnamefont{{Primack}}},
  \bibnamefont{and} \bibinfo{author}{\bibfnamefont{A.}~\bibnamefont{{Dekel}}},
  \bibinfo{journal}{\mnras} \textbf{\bibinfo{volume}{321}},
  \bibinfo{pages}{559} (\bibinfo{year}{2001}), \eprint{arXiv:astro-ph/9908159}.

\bibitem[{\citenamefont{{Gaisser} et~al.}(2001)\citenamefont{{Gaisser},
  {Honda}, {Lipari}, and {Stanev}}}]{2001ICRC....5.1643G}
\bibinfo{author}{\bibfnamefont{T.~K.} \bibnamefont{{Gaisser}}},
  \bibinfo{author}{\bibfnamefont{M.}~\bibnamefont{{Honda}}},
  \bibinfo{author}{\bibfnamefont{P.}~\bibnamefont{{Lipari}}}, \bibnamefont{and}
  \bibinfo{author}{\bibfnamefont{T.}~\bibnamefont{{Stanev}}}, in
  \emph{\bibinfo{booktitle}{International Cosmic Ray Conference}}
  (\bibinfo{year}{2001}), vol.~\bibinfo{volume}{5} of
  \emph{\bibinfo{series}{International Cosmic Ray Conference}}, pp.
  \bibinfo{pages}{1643--+}.

\bibitem[{\citenamefont{{Abdo} et~al.}(2009{\natexlab{a}})\citenamefont{{Abdo},
  {Ackermann}, {Ajello}, {Atwood}, {Axelsson}, {Baldini}, {Ballet},
  {Barbiellini}, {Bastieri}, {Battelino} et~al.}}]{2009PhRvL.102r1101A}
\bibinfo{author}{\bibfnamefont{A.~A.} \bibnamefont{{Abdo}}},
  \bibinfo{author}{\bibfnamefont{M.}~\bibnamefont{{Ackermann}}},
  \bibinfo{author}{\bibfnamefont{M.}~\bibnamefont{{Ajello}}},
  \bibinfo{author}{\bibfnamefont{W.~B.} \bibnamefont{{Atwood}}},
  \bibinfo{author}{\bibfnamefont{M.}~\bibnamefont{{Axelsson}}},
  \bibinfo{author}{\bibfnamefont{L.}~\bibnamefont{{Baldini}}},
  \bibinfo{author}{\bibfnamefont{J.}~\bibnamefont{{Ballet}}},
  \bibinfo{author}{\bibfnamefont{G.}~\bibnamefont{{Barbiellini}}},
  \bibinfo{author}{\bibfnamefont{D.}~\bibnamefont{{Bastieri}}},
  \bibinfo{author}{\bibfnamefont{M.}~\bibnamefont{{Battelino}}},
  \bibnamefont{et~al.}, \bibinfo{journal}{Physical Review Letters}
  \textbf{\bibinfo{volume}{102}}, \bibinfo{pages}{181101}
  (\bibinfo{year}{2009}{\natexlab{a}}), \eprint{0905.0025}.

\bibitem[{\citenamefont{{Aharonian} et~al.}(2008)\citenamefont{{Aharonian},
  {Akhperjanian}, {Barres de Almeida}, {Bazer-Bachi}, {Becherini}, {Behera},
  {Benbow}, {Bernl{\"o}hr}, {Boisson}, {Bochow} et~al.}}]{2008PhRvL.101z1104A}
\bibinfo{author}{\bibfnamefont{F.}~\bibnamefont{{Aharonian}}},
  \bibinfo{author}{\bibfnamefont{A.~G.} \bibnamefont{{Akhperjanian}}},
  \bibinfo{author}{\bibfnamefont{U.}~\bibnamefont{{Barres de Almeida}}},
  \bibinfo{author}{\bibfnamefont{A.~R.} \bibnamefont{{Bazer-Bachi}}},
  \bibinfo{author}{\bibfnamefont{Y.}~\bibnamefont{{Becherini}}},
  \bibinfo{author}{\bibfnamefont{B.}~\bibnamefont{{Behera}}},
  \bibinfo{author}{\bibfnamefont{W.}~\bibnamefont{{Benbow}}},
  \bibinfo{author}{\bibfnamefont{K.}~\bibnamefont{{Bernl{\"o}hr}}},
  \bibinfo{author}{\bibfnamefont{C.}~\bibnamefont{{Boisson}}},
  \bibinfo{author}{\bibfnamefont{A.}~\bibnamefont{{Bochow}}},
  \bibnamefont{et~al.}, \bibinfo{journal}{Physical Review Letters}
  \textbf{\bibinfo{volume}{101}}, \bibinfo{pages}{261104}
  (\bibinfo{year}{2008}), \eprint{0811.3894}.

\bibitem[{\citenamefont{{Aharonian}
  et~al.}(2009{\natexlab{a}})\citenamefont{{Aharonian}, {Akhperjanian},
  {Anton}, {Barres de Almeida}, {Bazer-Bachi}, {Becherini}, {Behera},
  {Bernl{\"o}hr}, {Bochow}, {Boisson} et~al.}}]{2009A&A...508..561A}
\bibinfo{author}{\bibfnamefont{F.}~\bibnamefont{{Aharonian}}},
  \bibinfo{author}{\bibfnamefont{A.~G.} \bibnamefont{{Akhperjanian}}},
  \bibinfo{author}{\bibfnamefont{G.}~\bibnamefont{{Anton}}},
  \bibinfo{author}{\bibfnamefont{U.}~\bibnamefont{{Barres de Almeida}}},
  \bibinfo{author}{\bibfnamefont{A.~R.} \bibnamefont{{Bazer-Bachi}}},
  \bibinfo{author}{\bibfnamefont{Y.}~\bibnamefont{{Becherini}}},
  \bibinfo{author}{\bibfnamefont{B.}~\bibnamefont{{Behera}}},
  \bibinfo{author}{\bibfnamefont{K.}~\bibnamefont{{Bernl{\"o}hr}}},
  \bibinfo{author}{\bibfnamefont{A.}~\bibnamefont{{Bochow}}},
  \bibinfo{author}{\bibfnamefont{C.}~\bibnamefont{{Boisson}}},
  \bibnamefont{et~al.}, \bibinfo{journal}{\aap} \textbf{\bibinfo{volume}{508}},
  \bibinfo{pages}{561} (\bibinfo{year}{2009}{\natexlab{a}}).

\bibitem[{\citenamefont{{Abdo} et~al.}(2010{\natexlab{a}})\citenamefont{{Abdo},
  {Ackermann}, {Ajello}, {Atwood}, {Baldini}, {Ballet}, {Barbiellini},
  {Bastieri}, {Baughman}, {Bechtol} et~al.}}]{2010PhRvL.104j1101A}
\bibinfo{author}{\bibfnamefont{A.~A.} \bibnamefont{{Abdo}}},
  \bibinfo{author}{\bibfnamefont{M.}~\bibnamefont{{Ackermann}}},
  \bibinfo{author}{\bibfnamefont{M.}~\bibnamefont{{Ajello}}},
  \bibinfo{author}{\bibfnamefont{W.~B.} \bibnamefont{{Atwood}}},
  \bibinfo{author}{\bibfnamefont{L.}~\bibnamefont{{Baldini}}},
  \bibinfo{author}{\bibfnamefont{J.}~\bibnamefont{{Ballet}}},
  \bibinfo{author}{\bibfnamefont{G.}~\bibnamefont{{Barbiellini}}},
  \bibinfo{author}{\bibfnamefont{D.}~\bibnamefont{{Bastieri}}},
  \bibinfo{author}{\bibfnamefont{B.~M.} \bibnamefont{{Baughman}}},
  \bibinfo{author}{\bibfnamefont{K.}~\bibnamefont{{Bechtol}}},
  \bibnamefont{et~al.}, \bibinfo{journal}{Physical Review Letters}
  \textbf{\bibinfo{volume}{104}}, \bibinfo{pages}{101101}
  (\bibinfo{year}{2010}{\natexlab{a}}), \eprint{1002.3603}.

\bibitem[{\citenamefont{{Sreekumar} et~al.}(1998)\citenamefont{{Sreekumar},
  {Bertsch}, {Dingus}, {Esposito}, {Fichtel}, {Hartman}, {Hunter}, {Kanbach},
  {Kniffen}, {Lin} et~al.}}]{1998ApJ...494..523S}
\bibinfo{author}{\bibfnamefont{P.}~\bibnamefont{{Sreekumar}}},
  \bibinfo{author}{\bibfnamefont{D.~L.} \bibnamefont{{Bertsch}}},
  \bibinfo{author}{\bibfnamefont{B.~L.} \bibnamefont{{Dingus}}},
  \bibinfo{author}{\bibfnamefont{J.~A.} \bibnamefont{{Esposito}}},
  \bibinfo{author}{\bibfnamefont{C.~E.} \bibnamefont{{Fichtel}}},
  \bibinfo{author}{\bibfnamefont{R.~C.} \bibnamefont{{Hartman}}},
  \bibinfo{author}{\bibfnamefont{S.~D.} \bibnamefont{{Hunter}}},
  \bibinfo{author}{\bibfnamefont{G.}~\bibnamefont{{Kanbach}}},
  \bibinfo{author}{\bibfnamefont{D.~A.} \bibnamefont{{Kniffen}}},
  \bibinfo{author}{\bibfnamefont{Y.~C.} \bibnamefont{{Lin}}},
  \bibnamefont{et~al.}, \bibinfo{journal}{\apj} \textbf{\bibinfo{volume}{494}},
  \bibinfo{pages}{523} (\bibinfo{year}{1998}), \eprint{arXiv:astro-ph/9709257}.

\bibitem[{\citenamefont{{Abdo} et~al.}(2009{\natexlab{b}})\citenamefont{{Abdo},
  {Ackermann}, {Ajello}, {Atwood}, {Axelsson}, {Baldini}, {Ballet},
  {Barbiellini}, {Bastieri}, {Baughman} et~al.}}]{2009ApJ...703.1249A}
\bibinfo{author}{\bibfnamefont{A.~A.} \bibnamefont{{Abdo}}},
  \bibinfo{author}{\bibfnamefont{M.}~\bibnamefont{{Ackermann}}},
  \bibinfo{author}{\bibfnamefont{M.}~\bibnamefont{{Ajello}}},
  \bibinfo{author}{\bibfnamefont{W.~B.} \bibnamefont{{Atwood}}},
  \bibinfo{author}{\bibfnamefont{M.}~\bibnamefont{{Axelsson}}},
  \bibinfo{author}{\bibfnamefont{L.}~\bibnamefont{{Baldini}}},
  \bibinfo{author}{\bibfnamefont{J.}~\bibnamefont{{Ballet}}},
  \bibinfo{author}{\bibfnamefont{G.}~\bibnamefont{{Barbiellini}}},
  \bibinfo{author}{\bibfnamefont{D.}~\bibnamefont{{Bastieri}}},
  \bibinfo{author}{\bibfnamefont{B.~M.} \bibnamefont{{Baughman}}},
  \bibnamefont{et~al.}, \bibinfo{journal}{\apj} \textbf{\bibinfo{volume}{703}},
  \bibinfo{pages}{1249} (\bibinfo{year}{2009}{\natexlab{b}}),
  \eprint{0908.1171}.

\bibitem[{\citenamefont{{Abdo} et~al.}(2009{\natexlab{c}})\citenamefont{{Abdo},
  {Ackermann}, {Ajello}, {Anderson}, {Atwood}, {Axelsson}, {Baldini}, {Ballet},
  {Barbiellini}, {Bastieri} et~al.}}]{2009PhRvL.103y1101A}
\bibinfo{author}{\bibfnamefont{A.~A.} \bibnamefont{{Abdo}}},
  \bibinfo{author}{\bibfnamefont{M.}~\bibnamefont{{Ackermann}}},
  \bibinfo{author}{\bibfnamefont{M.}~\bibnamefont{{Ajello}}},
  \bibinfo{author}{\bibfnamefont{B.}~\bibnamefont{{Anderson}}},
  \bibinfo{author}{\bibfnamefont{W.~B.} \bibnamefont{{Atwood}}},
  \bibinfo{author}{\bibfnamefont{M.}~\bibnamefont{{Axelsson}}},
  \bibinfo{author}{\bibfnamefont{L.}~\bibnamefont{{Baldini}}},
  \bibinfo{author}{\bibfnamefont{J.}~\bibnamefont{{Ballet}}},
  \bibinfo{author}{\bibfnamefont{G.}~\bibnamefont{{Barbiellini}}},
  \bibinfo{author}{\bibfnamefont{D.}~\bibnamefont{{Bastieri}}},
  \bibnamefont{et~al.}, \bibinfo{journal}{Physical Review Letters}
  \textbf{\bibinfo{volume}{103}}, \bibinfo{pages}{251101}
  (\bibinfo{year}{2009}{\natexlab{c}}), \eprint{0912.0973}.

\bibitem[{\citenamefont{{Hunter} et~al.}(1997)\citenamefont{{Hunter},
  {Bertsch}, {Catelli}, {Dame}, {Digel}, {Dingus}, {Esposito}, {Fichtel},
  {Hartman}, {Kanbach} et~al.}}]{1997ApJ...481..205H}
\bibinfo{author}{\bibfnamefont{S.~D.} \bibnamefont{{Hunter}}},
  \bibinfo{author}{\bibfnamefont{D.~L.} \bibnamefont{{Bertsch}}},
  \bibinfo{author}{\bibfnamefont{J.~R.} \bibnamefont{{Catelli}}},
  \bibinfo{author}{\bibfnamefont{T.~M.} \bibnamefont{{Dame}}},
  \bibinfo{author}{\bibfnamefont{S.~W.} \bibnamefont{{Digel}}},
  \bibinfo{author}{\bibfnamefont{B.~L.} \bibnamefont{{Dingus}}},
  \bibinfo{author}{\bibfnamefont{J.~A.} \bibnamefont{{Esposito}}},
  \bibinfo{author}{\bibfnamefont{C.~E.} \bibnamefont{{Fichtel}}},
  \bibinfo{author}{\bibfnamefont{R.~C.} \bibnamefont{{Hartman}}},
  \bibinfo{author}{\bibfnamefont{G.}~\bibnamefont{{Kanbach}}},
  \bibnamefont{et~al.}, \bibinfo{journal}{\apj} \textbf{\bibinfo{volume}{481}},
  \bibinfo{pages}{205} (\bibinfo{year}{1997}).

\bibitem[{\citenamefont{{Bergstr{\"o}m}
  et~al.}(1998)\citenamefont{{Bergstr{\"o}m}, {Ullio}, and
  {Buckley}}}]{1998APh.....9..137B}
\bibinfo{author}{\bibfnamefont{L.}~\bibnamefont{{Bergstr{\"o}m}}},
  \bibinfo{author}{\bibfnamefont{P.}~\bibnamefont{{Ullio}}}, \bibnamefont{and}
  \bibinfo{author}{\bibfnamefont{J.~H.} \bibnamefont{{Buckley}}},
  \bibinfo{journal}{Astroparticle Physics} \textbf{\bibinfo{volume}{9}},
  \bibinfo{pages}{137} (\bibinfo{year}{1998}), \eprint{arXiv:astro-ph/9712318}.

\bibitem[{\citenamefont{{Aharonian}
  et~al.}(2006{\natexlab{a}})\citenamefont{{Aharonian}, {Akhperjanian},
  {Bazer-Bachi}, {Beilicke}, {Benbow}, {Berge}, {Bernl{\"o}hr}, {Boisson},
  {Bolz}, {Borrel} et~al.}}]{2006Natur.439..695A}
\bibinfo{author}{\bibfnamefont{F.}~\bibnamefont{{Aharonian}}},
  \bibinfo{author}{\bibfnamefont{A.~G.} \bibnamefont{{Akhperjanian}}},
  \bibinfo{author}{\bibfnamefont{A.~R.} \bibnamefont{{Bazer-Bachi}}},
  \bibinfo{author}{\bibfnamefont{M.}~\bibnamefont{{Beilicke}}},
  \bibinfo{author}{\bibfnamefont{W.}~\bibnamefont{{Benbow}}},
  \bibinfo{author}{\bibfnamefont{D.}~\bibnamefont{{Berge}}},
  \bibinfo{author}{\bibfnamefont{K.}~\bibnamefont{{Bernl{\"o}hr}}},
  \bibinfo{author}{\bibfnamefont{C.}~\bibnamefont{{Boisson}}},
  \bibinfo{author}{\bibfnamefont{O.}~\bibnamefont{{Bolz}}},
  \bibinfo{author}{\bibfnamefont{V.}~\bibnamefont{{Borrel}}},
  \bibnamefont{et~al.}, \bibinfo{journal}{\nat} \textbf{\bibinfo{volume}{439}},
  \bibinfo{pages}{695} (\bibinfo{year}{2006}{\natexlab{a}}),
  \eprint{arXiv:astro-ph/0603021}.

\bibitem[{\citenamefont{{Mayer-Hasselwander}
  et~al.}(1998)\citenamefont{{Mayer-Hasselwander}, {Bertsch}, {Dingus},
  {Eckart}, {Esposito}, {Genzel}, {Hartman}, {Hunter}, {Kanbach}, {Kniffen}
  et~al.}}]{1998A&A...335..161M}
\bibinfo{author}{\bibfnamefont{H.~A.} \bibnamefont{{Mayer-Hasselwander}}},
  \bibinfo{author}{\bibfnamefont{D.~L.} \bibnamefont{{Bertsch}}},
  \bibinfo{author}{\bibfnamefont{B.~L.} \bibnamefont{{Dingus}}},
  \bibinfo{author}{\bibfnamefont{A.}~\bibnamefont{{Eckart}}},
  \bibinfo{author}{\bibfnamefont{J.~A.} \bibnamefont{{Esposito}}},
  \bibinfo{author}{\bibfnamefont{R.}~\bibnamefont{{Genzel}}},
  \bibinfo{author}{\bibfnamefont{R.~C.} \bibnamefont{{Hartman}}},
  \bibinfo{author}{\bibfnamefont{S.~D.} \bibnamefont{{Hunter}}},
  \bibinfo{author}{\bibfnamefont{G.}~\bibnamefont{{Kanbach}}},
  \bibinfo{author}{\bibfnamefont{D.~A.} \bibnamefont{{Kniffen}}},
  \bibnamefont{et~al.}, \bibinfo{journal}{\aap} \textbf{\bibinfo{volume}{335}},
  \bibinfo{pages}{161} (\bibinfo{year}{1998}).

\bibitem[{\citenamefont{{Aharonian}
  et~al.}(2006{\natexlab{b}})\citenamefont{{Aharonian}, {Akhperjanian},
  {Bazer-Bachi}, {Beilicke}, {Benbow}, {Berge}, {Bernl{\"o}hr}, {Boisson},
  {Bolz}, {Borrel} et~al.}}]{2006PhRvL..97v1102A}
\bibinfo{author}{\bibfnamefont{F.}~\bibnamefont{{Aharonian}}},
  \bibinfo{author}{\bibfnamefont{A.~G.} \bibnamefont{{Akhperjanian}}},
  \bibinfo{author}{\bibfnamefont{A.~R.} \bibnamefont{{Bazer-Bachi}}},
  \bibinfo{author}{\bibfnamefont{M.}~\bibnamefont{{Beilicke}}},
  \bibinfo{author}{\bibfnamefont{W.}~\bibnamefont{{Benbow}}},
  \bibinfo{author}{\bibfnamefont{D.}~\bibnamefont{{Berge}}},
  \bibinfo{author}{\bibfnamefont{K.}~\bibnamefont{{Bernl{\"o}hr}}},
  \bibinfo{author}{\bibfnamefont{C.}~\bibnamefont{{Boisson}}},
  \bibinfo{author}{\bibfnamefont{O.}~\bibnamefont{{Bolz}}},
  \bibinfo{author}{\bibfnamefont{V.}~\bibnamefont{{Borrel}}},
  \bibnamefont{et~al.}, \bibinfo{journal}{Physical Review Letters}
  \textbf{\bibinfo{volume}{97}}, \bibinfo{pages}{221102}
  (\bibinfo{year}{2006}{\natexlab{b}}), \eprint{arXiv:astro-ph/0610509}.

\bibitem[{\citenamefont{{Abdo} et~al.}(2009{\natexlab{d}})\citenamefont{{Abdo},
  {Ackermann}, {Ajello}, {Atwood}, {Axelsson}, {Baldini}, {Ballet}, {Band},
  {Barbiellini}, {Bastieri} et~al.}}]{2009ApJS..183...46A}
\bibinfo{author}{\bibfnamefont{A.~A.} \bibnamefont{{Abdo}}},
  \bibinfo{author}{\bibfnamefont{M.}~\bibnamefont{{Ackermann}}},
  \bibinfo{author}{\bibfnamefont{M.}~\bibnamefont{{Ajello}}},
  \bibinfo{author}{\bibfnamefont{W.~B.} \bibnamefont{{Atwood}}},
  \bibinfo{author}{\bibfnamefont{M.}~\bibnamefont{{Axelsson}}},
  \bibinfo{author}{\bibfnamefont{L.}~\bibnamefont{{Baldini}}},
  \bibinfo{author}{\bibfnamefont{J.}~\bibnamefont{{Ballet}}},
  \bibinfo{author}{\bibfnamefont{D.~L.} \bibnamefont{{Band}}},
  \bibinfo{author}{\bibfnamefont{G.}~\bibnamefont{{Barbiellini}}},
  \bibinfo{author}{\bibfnamefont{D.}~\bibnamefont{{Bastieri}}},
  \bibnamefont{et~al.}, \bibinfo{journal}{\apjs}
  \textbf{\bibinfo{volume}{183}}, \bibinfo{pages}{46}
  (\bibinfo{year}{2009}{\natexlab{d}}), \eprint{0902.1340}.

\bibitem[{\citenamefont{{Aharonian}
  et~al.}(2009{\natexlab{b}})\citenamefont{{Aharonian}, {Akhperjanian},
  {Anton}, {Barres de Almeida}, {Bazer-Bachi}, {Becherini}, {Behera},
  {Bernl{\"o}hr}, {Boisson}, {Bochow} et~al.}}]{2009A&A...503..817A}
\bibinfo{author}{\bibfnamefont{F.}~\bibnamefont{{Aharonian}}},
  \bibinfo{author}{\bibfnamefont{A.~G.} \bibnamefont{{Akhperjanian}}},
  \bibinfo{author}{\bibfnamefont{G.}~\bibnamefont{{Anton}}},
  \bibinfo{author}{\bibfnamefont{U.}~\bibnamefont{{Barres de Almeida}}},
  \bibinfo{author}{\bibfnamefont{A.~R.} \bibnamefont{{Bazer-Bachi}}},
  \bibinfo{author}{\bibfnamefont{Y.}~\bibnamefont{{Becherini}}},
  \bibinfo{author}{\bibfnamefont{B.}~\bibnamefont{{Behera}}},
  \bibinfo{author}{\bibfnamefont{K.}~\bibnamefont{{Bernl{\"o}hr}}},
  \bibinfo{author}{\bibfnamefont{C.}~\bibnamefont{{Boisson}}},
  \bibinfo{author}{\bibfnamefont{A.}~\bibnamefont{{Bochow}}},
  \bibnamefont{et~al.}, \bibinfo{journal}{\aap} \textbf{\bibinfo{volume}{503}},
  \bibinfo{pages}{817} (\bibinfo{year}{2009}{\natexlab{b}}),
  \eprint{0906.1247}.

\bibitem[{\citenamefont{{Abdo} et~al.}(2010{\natexlab{b}})\citenamefont{{Abdo},
  {Ackermann}, {Ajello}, {Atwood}, {Baldini}, {Ballet}, {Barbiellini},
  {Bastieri}, {Bechtol}, {Bellazzini} et~al.}}]{2010PhRvL.104i1302A}
\bibinfo{author}{\bibfnamefont{A.~A.} \bibnamefont{{Abdo}}},
  \bibinfo{author}{\bibfnamefont{M.}~\bibnamefont{{Ackermann}}},
  \bibinfo{author}{\bibfnamefont{M.}~\bibnamefont{{Ajello}}},
  \bibinfo{author}{\bibfnamefont{W.~B.} \bibnamefont{{Atwood}}},
  \bibinfo{author}{\bibfnamefont{L.}~\bibnamefont{{Baldini}}},
  \bibinfo{author}{\bibfnamefont{J.}~\bibnamefont{{Ballet}}},
  \bibinfo{author}{\bibfnamefont{G.}~\bibnamefont{{Barbiellini}}},
  \bibinfo{author}{\bibfnamefont{D.}~\bibnamefont{{Bastieri}}},
  \bibinfo{author}{\bibfnamefont{K.}~\bibnamefont{{Bechtol}}},
  \bibinfo{author}{\bibfnamefont{R.}~\bibnamefont{{Bellazzini}}},
  \bibnamefont{et~al.}, \bibinfo{journal}{Physical Review Letters}
  \textbf{\bibinfo{volume}{104}}, \bibinfo{pages}{091302}
  (\bibinfo{year}{2010}{\natexlab{b}}), \eprint{1001.4836}.

\end{thebibliography}

\end{document}